\documentclass{article}

\usepackage{arxiv}

\usepackage[utf8]{inputenc} % allow utf-8 input
\usepackage[T1]{fontenc}    % use 8-bit T1 fonts
\usepackage{hyperref}       % hyperlinks
\usepackage{url}            % simple URL typesetting
\usepackage{booktabs}       % professional-quality tables
\usepackage{amsfonts}       % blackboard math symbols
\usepackage{nicefrac}       % compact symbols for 1/2, etc.
\usepackage{microtype}      % microtypography
\usepackage{lipsum}     % Can be removed after putting your text content
\usepackage{graphicx}
\usepackage{doi}

\usepackage{amsmath}
\usepackage{amssymb}
\usepackage{latexsym}
\usepackage{graphicx}
\usepackage{natbib}
\usepackage{color}
\bibpunct{(}{)}{;}{a}{}{,}

\usepackage{amstext}

\newcommand{\arcsec}{\ensuremath{^{\prime\prime}}}
\newcommand{\apj}{ApJ}
\newcommand{\aap}{A\&A}
\newcommand{\apjl}{ApJL}
\newcommand{\apjs}{ApJS}
\newcommand{\nat}{Nature}
\newcommand{\solphys}{Sol. Phys.}

\title{The fine-scale structure of polar coronal holes}

\author{ \href{https://orcid.org/0000-0001-5678-9002}{\includegraphics[scale=0.06]{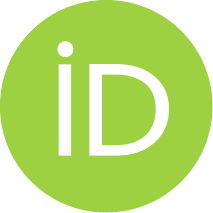}\hspace{1mm}R. J. Morton}\\
Department of Maths Physics and Electrical Engineering \\
Northumbria University, UK \\
\texttt{richard.morton@northumbria.ac.uk}\\
\And
R. Cunningham \\
Department of Maths Physics and Electrical Engineering \\
Northumbria University, UK}

\begin{document}
\maketitle

\begin{abstract}
Coronal holes are thought to be composed of relatively broad columnar structures known as plumes. Here we demonstrate that the plumes (and inter-plumes) in polar coronal holes are composed of fine-scale filamentary
structure, with average scales of 2-10$^{\arcsec}$. The fine structure is the off-limb analogue of the previously found 'plumelets' of \cite{Uritsky_2021}. The off-limb observations enable an examination of the fine-structure without the influence of the underlying atmosphere along the line of sight. Hence, we show that the fine-scale structure is present at least until the edge of the field of view of the Solar Dynamics Observatory. The fine structure is found to have spatial distribution that follows a 
$k^{-1}$ power law perpendicular to the inferred magnetic field direction. For a small sample of the fine structure, the cross-sectional profiles are measured as a function of height. In some cases, the measurements indicate that the fine structure  expands super-radially, consistent with existing models of polar field expansion and the expansion of the plumes. We discuss the
implications of the presence of the fine structure with respect to understanding wave propagation in the coronal holes and their contribution to powering the solar wind.
\end{abstract}

\keywords{The Sun (1693), Solar corona (1483), Solar coronal holes (1484), Solar coronal plumes (2039)}

%\maketitle

\section{Introduction}\label{sec1}

Plumes, observed in both polar regions and lower latitude coronal holes, are a common coronal feature that are rooted in near-unipolar fields comprising the magnetic network \citep{Newkirk_1968,Wang_1995,DeForest_1997}. The plumes appear after the emergence of ephemeral regions and can live for several days \citep{Wang_2008}. Plumes generally appear as diffuse, columnar features in EUV images, often termed `beam' plumes. The presence of an additional plume structure, the `curtain' plume, has also been suggested in the past to explain the (false) impression in EUV images that plumes are mainly rooted on the far side of the Sun \citep{Wilhelm_2011}. The plumes are found to extend up to $\sim40$~R$_\odot$ in the polar regions, expanding super-radially \citep{Deforest_2001}, hence link many regions of the Sun's outer atmosphere. Over the decades, they have received significant attention as they likely play an important role in supplying mass and energy into the solar wind.

Plumes exhibit a variety of dynamic behaviour. Outflows have been observed throughout the coronal holes (in both plume and inter-plume regions), seemingly starting in the transition region \citep{TIAetal2010} and increasing through the low corona \citep{Rifai_Habbal_1995,Teriaca_2003,MORetal2015}. The plumes also display outwardly-moving, quasi-periodic disturbances visible as intensity fluctuations in imagers. It remains unclear whether these events are slow magnetoacoustic waves or mass flows \citep[e.g.,][]{DeForest_1998,BANetal2009,McIntosh_2010,Tian_2011,Krishna_Prasad_2011,Pucci_2014}.
\cite{McIntosh_2010} suggested the occurrence of the fluctuations is connected to jet activity in the upper chromosphere, and is responsible for mass loading heated plasma into the plumes. \cite{Samanta_2015} provided evidence that some of these fluctuations are indeed initiated by spicules. A potentially related phenomenon are small-scale coronal jets discussed by \cite{Raouafi_2014}, although they seem to be on a larger scale than spicules. \cite{Raouafi_2014} suggest that the coronal jets provide mass and energy to initiate and sustain plumes. 

Plumes in polar coronal holes also demonstrate quasi-periodic transverse displacements in intensity images \citep{THUetal2014}, which are associated with intensity features at smaller scales. These motions have been interpreted as the kink mode, which is essentially a surface Alfv\'en wave \citep[or Alfv\'enic waves, ][]{GOOetal2012}. The Doppler velocities of coronal emission lines from polar regions also demonstrate the corresponding signal of these waves \citep{MORetal2015}. Given the omni-presence of Alfv\'enic waves throughout the corona \citep{MCIetal2011,MORetal2019} it should be expected that the Alfv\'enic waves are also present in equatorial coronal hole plumes, however, there is currently no specific confirmation of this. The Alfv\'enic fluctuations are regular and present across the coronal holes, plumes and interplumes, indicating that the main driver is convective buffeting and/or the mode conversions of \textit{p}-modes. Energy estimates suggest the observed fluctuations carry enough energy to provide a significant contribution to the heating or acceleration requirements of the solar wind \citep{MCIetal2011,MORetal2015,WEBetal2018}. Although not all Alfv\'enic modes, e.g., rotational motions, are directly observable in the corona (yet) and so the actual energy flux of Alfv\'enic waves in coronal holes is unknown.     

\medskip

It has recently been recognised that plumes contain fine-structure. The fine-structure was first noted in \cite{DeForest_1997}, who observed transient intensity enhancements at scales of 10\arcsec. It was proposed by \cite{Gabriel_2009} that, in order to explain the appearance and behaviour of curtain plumes, they must be composed of fine-scale structures which they termed micro-plumes (or network plumes). \cite{Wilhelm_2011} suggest that `beam' plumes may also be composed of micro-plumes. Plumes generally appear as diffuse, columnar features in EUV images. However, image processing reveals that plumes in equatorial coronal holes are composed of filamentary substructure \citep{Uritsky_2021} \citep[also partialy visible in the processed images of][]{Raouafi_2014}. They were given the name \textit{`plumelets'} and found to vary on the time-scales of minutes. The network regions below plumes are known to consist of small-scale `network' jets \citep{Tian_2014,Qi_2019}. \citet{Kumar_2022} extended the investigation of the substructure, finding variations on the order of 5 minutes, in accordance with the previous studies that had examined the quasi-periodic disturbances integrated over the diffuse plumes. \citet{Kumar_2022} connected the appearance of the substructure with the occurrence of coronal jets at the base of the plumes, and suggested that small-scale quasi-periodic reconnection is responsible for producing mass-flows that feed the substructure and ultimately contribute to mass loss via the solar wind. Recently, \cite{Raouafi_2023} demonstrated the phenomenon is widespread in coronal holes and suggest the coronal jets (or jetlets as the authors call them) are the primary source of the heating and driving of the solar wind.

\medskip

In this work, we examine the spatial scales of the fine-scale structure within polar coronal holes. We demonstrate that both plume and inter-plume regions show the presence of fine-scale structure. This filamentary structure is superimposed upon a bright, diffuse background emission and is only revealed clearly after removing the diffuse component.
The emission from the fine-scale plasma structure is weak, representing only $\sim1\%$ of the total emission locally (comparable to the photon noise within the images). Given the substantive role of plasma density in EUV spectral line brightness, it would suggest that the coronal holes are littered with weakly over-dense striations. In Section~\ref{sec:data_prep} we discuss the processing of the data in order to best reveal the weak emission. Then in Section~\ref{sec:results_perp} we highlight the perpendicular scales associated with the emission structures and Section~\ref{sec:results_height} discuses variations with height. In Section~\ref{sec:discussion} we discuss the results and implications, undertaking a relatively simple simulation to examine whether the inherent integration of emission over the optically thin corona skews the measurable spatial distribution of emitting fine structures.

\section{Observations and data preparation}\label{sec:data_prep}

The data used for this study were taken with the Atmospheric Imaging Assembly \citep[AIA][]{LEMetal2012} onboard the Solar Dynamics Observatory \citep[SDO][]{PESetal2012}. The date of the observations is the 30 January 2011 between 19:00~UT and 21:00~UT. A polar coronal hole located at the south pole of the Sun is the primary target for the following analysis.

\begin{figure}
\centering
    \includegraphics[trim=25 400 300px 30px, clip, scale=0.92]{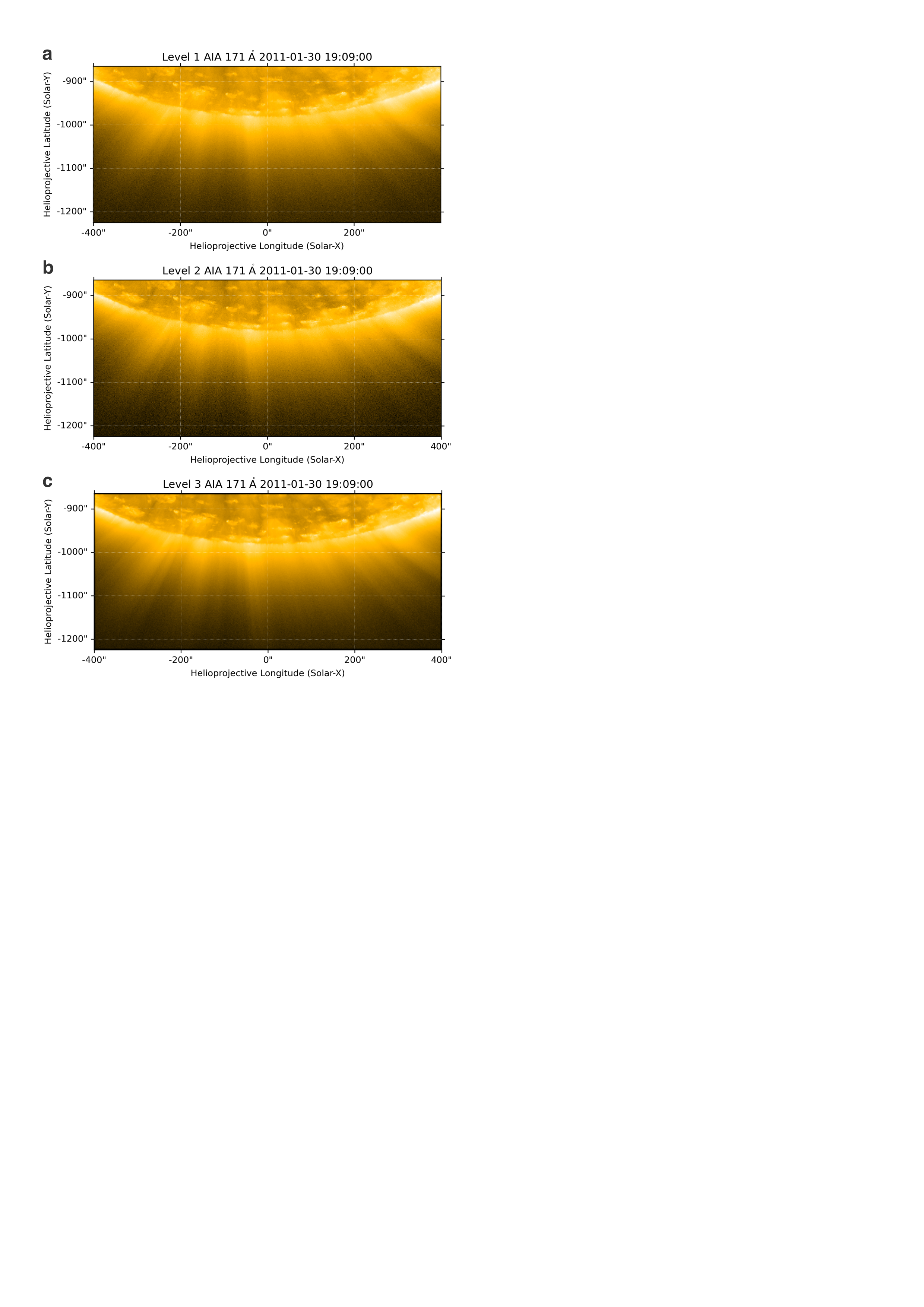}
    \caption{Polar coronal hole observed the 30 Jan 2011. The panels show the different
    stages of processing applied to the data. Panel a is the original Level 1 data. Panel b is
    the deconvolved data (Level 2). Panel c is the noise-gated data (Level 3). The plumes are substantially clearer in the Level 3 data, highlighting the removal of scattered light and
    significant noise reduction.}\label{fig:poles}
\end{figure}

A number of processing steps are required in order to examine the faintest features in the polar regions. The Level 1 data is not processed to the standard Level 1.5 data, typically performed with an \textit{AIA$\_$prep} routine. The
standard data preparation rotates the data and interpolates the images down to a specified plate scale. This procedure is known to introduce artefacts into the data that only becomes apparent in regions with low signal \citep{cheung}. We intend to examine the nature of faint features, hence need to minimise additional sources of noise. A cut-out of a Level 1 file focused on the polar region is displayed in Figure~\ref{fig:poles}a. The coronal hole consists of large-scale ($> 30^{\prime\prime}$) plumes, consistent with earlier observations \cite[e.g.,][]{DeForest_1997}. If one looks closely enough, you can get the impression
that smaller-scale features are present within the plumes. The image also reveals that the emission in the coronal hole is faint and becomes dominated by noise at a reasonably low height ($\sim$1.15~$R_\odot$). 

Scattered light from the optical elements in the telescope is also present in the images. To remove the scattered light, we deconvolve the data using Richardson-Lucy deconvolution and the point spread function determined in \cite{grigis_2012}. This is called Level 2 data. An example of the deconvolved cut-out region is given in 
Figure~\ref{fig:poles}b. It can be seen that there is an overall decrease in the level of emission due to removal of the scattered light, however the signal-to-noise (S/N) has also decreased. In order to improve the S/N levels, we employ the noise-gating technique developed in \cite{DeForest_2017}. The method uses locally adaptive filters in the Fourier domain to remove background noise in local image neighbourhoods. We use the gating method on $12\times12\times12$ $(nx,ny,nt)$ cubes with the threshold
factor set to 3. The noise-gating is applied to the cut-out region shown in Figure~\ref{fig:poles}b. The result 
of the filtering is shown in Figure~\ref{fig:poles}c and called Level 3 data.

\begin{figure}
    \includegraphics[scale=0.55]{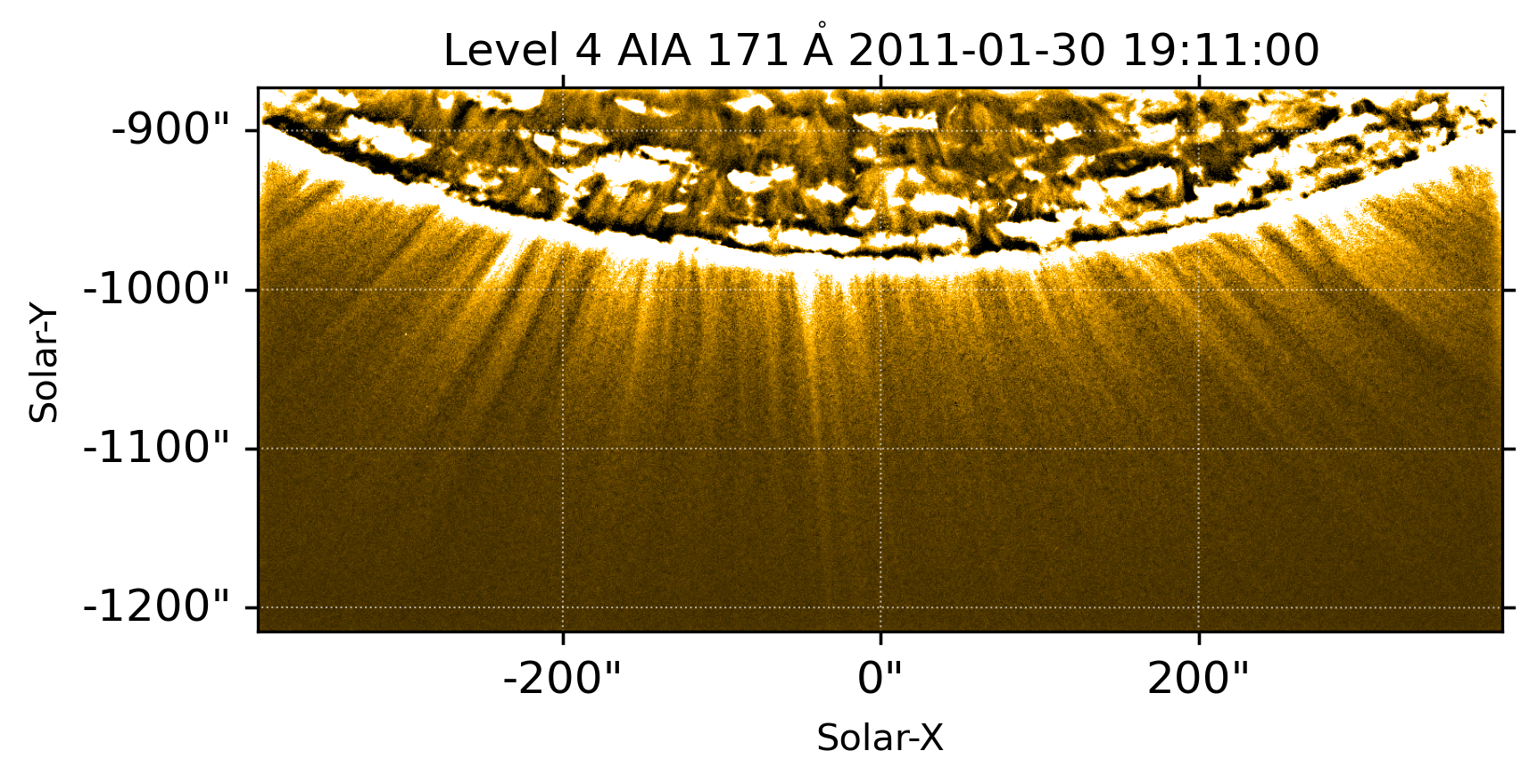}
    \includegraphics[scale=0.55]{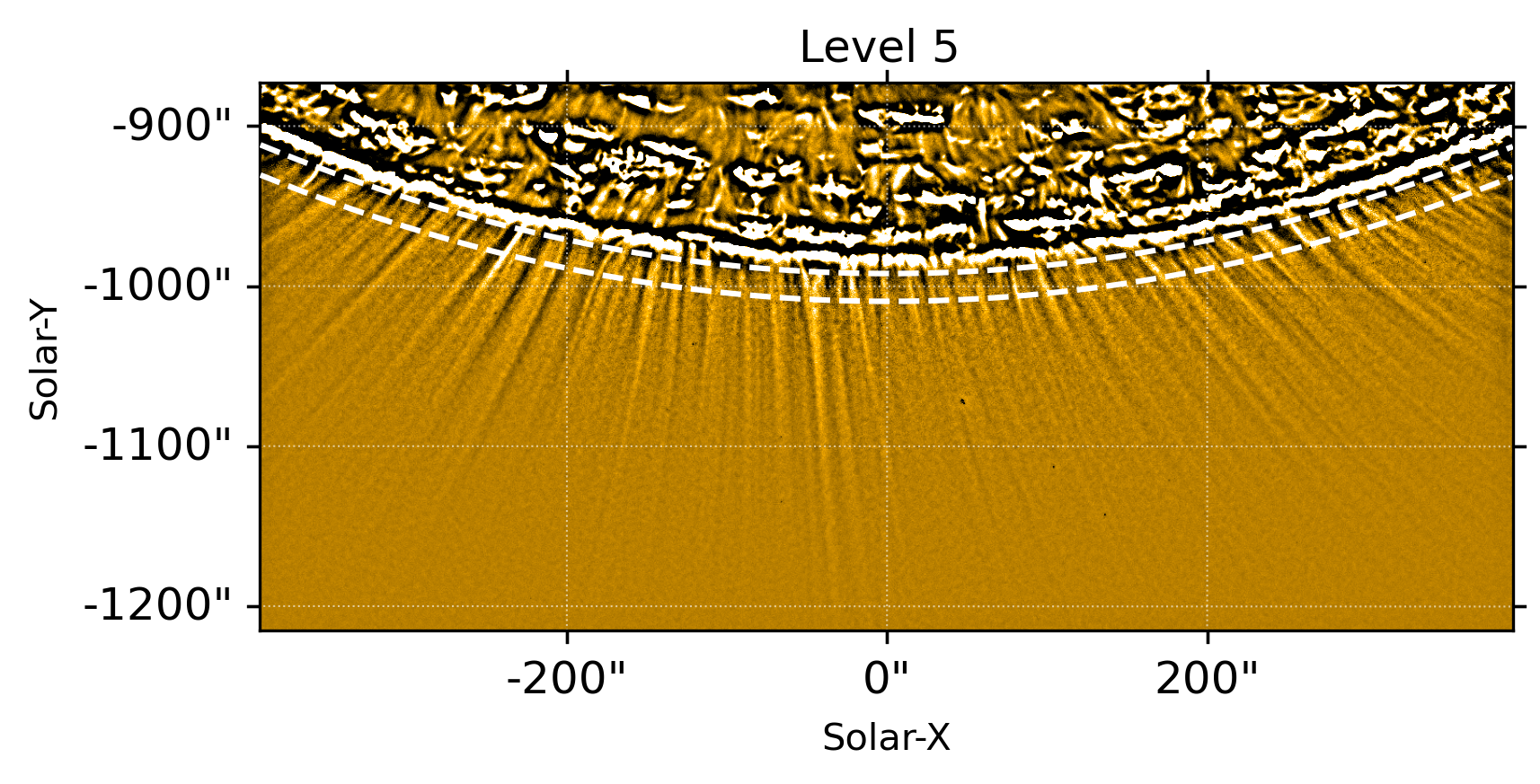}
    \caption{The polar region subject to further processing. The top panel displays
    the Level 4 \textit{minsmoothed} data. The lower panel shows the unsharp masked 
    Level 5 data, which is generated from a temporal average over $420$~s. The white dashed lines
    show the region where the curves are extracted for visual and Fourier analysis.}
    \label{fig:usm_pole}
\end{figure}

The Level 3 images now reveal new features (\textbf{in the Appendix we estimate the reduction in the noise levels from the processing steps}). The first is that the plume structures are now are easily followed into the higher parts of the atmosphere within the image, and are distinct until the edge of the field of view. And secondly, the finer-scale, filamentary structure across these large plumes is also more evident. The fine-scale structures can be seen more easily by removing the large-scale background emission. The \textit{minsmoothing} technique is used to filter the image \citep{DeForest_2015}, which is carried out by applying an erosion operator with circular kernel of radius $3^{\arcsec}$ and convolving the eroded image with a two dimensional Gaussian with $\sigma=6^{\prime\prime}$ . The smooth background is subtracted from the original data to produce the Level 4 data, which is shown in Figure~\ref{fig:usm_pole} upper panel. The fine-scale structure within the coronal hole now stands out clearly. 

\section{Results}
\subsection{Structure of plumes}\label{sec:results_perp}
The Level 4 data shows that the larger plumes, and inter-plume regions, are composed of structures that have widths close to the resolution limit of SDO/AIA (Figure~\ref{fig:usm_pole}). The plumes appear to be composed of bright small-scale structures, which are visible at the higher heights. In contrast, the inter-plume regions show dimmer structures that are less visible at the same heights. Previous work by \cite{THUetal2014,Weberg_2020} had also shown the presence of
fine-structure in polar coronal holes, demonstrating that it supports transverse wave motions; although the authors did not comment on the fine-structure.

As mentioned in the introduction, similar fine-scale structure has been observed in
plumes found in equatorial coronal holes by \cite{Uritsky_2021}. Although, the report of \cite{Uritsky_2021} states the filamentary structure is found at the base of coronal plumes and apparently only visible on length-scales of $<150^{\arcsec}$ along the magnetic field orientation. In the polar images shown here it is difficult to locate the regions where the fine-structure originates, likely due to some features being rooted on the far side of the pole. However, it is seen that the fine-structure is present all the way to the edge of the SDO FOV, suggesting the plumes show density enhancements up to $\sim300^{\prime\prime}$ above the limb. The reason for the enhanced visibility of the fine-scale features in the polar regions is potentially due to the lack of overlap with strongly emitting features along the line-of-sight, which inevitably happens in on-disk observations.  

\begin{figure*}
    \includegraphics[scale=0.55]{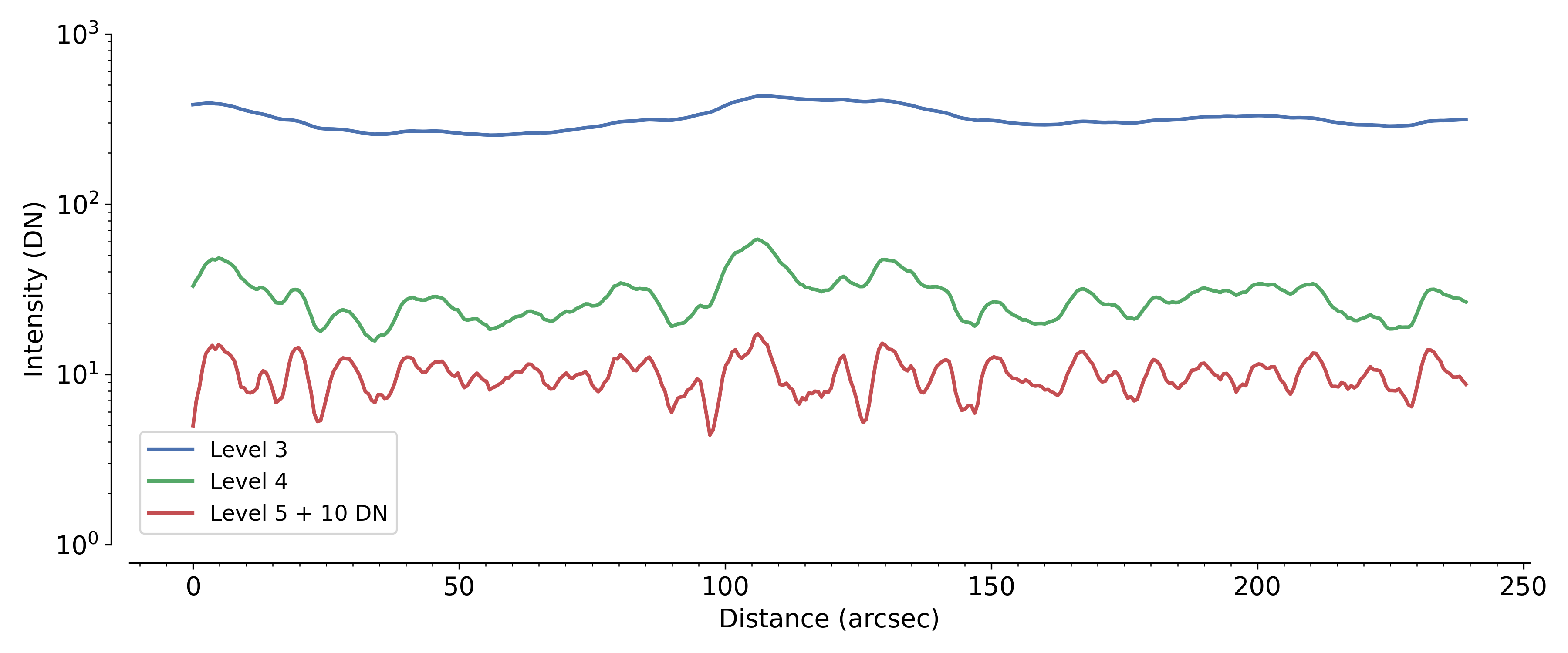}
    \caption{Visual examination of intensity structures. The different curves correspond to
    the intensity in polar coronal hole across the plumes at a fixed radial distance from the photosphere. The intensity is a spatial and temporal average. For visibility we have added a constant value to the Level 5 curve (10~DN). It can be observed that the
    Level 4 and 5 processing steps enhance the fine-scale structures.}
    \label{fig:int_curves}
\end{figure*}

We can estimate the characteristic spatial scales transverse to the
filamentation. Unsharp masking is employed to further isolate the fine-scale structure, with the Level 4 images subject to a smoothing using a Gaussian kernel with a $\sigma=3^{\prime\prime}$. The smooth images are subtracted from the original Level 4 image, leaving only the fine-scale features 
(Figure~\ref{fig:usm_pole} lower panel). To provide a comparison between the scales of structures visible in each of the processing steps,
we extract 30 arcs at set radial distances from the photosphere, between 1.017~$R_\odot$ and 1.036~$R_\odot$ (demarcated by the curves in Figure~\ref{fig:usm_pole} lower panel) for 420~s of data. The temporal and radial averages of these curves are shown in Figure~\ref{fig:int_curves}. There difference in structure visibility between the Level 3 and Level 5 data is clear. The Level 3 data is dominated by the background emission, with plumes superimposed on the top of this. The Level 4 data reveals the plumes more clearly, possessing intensity features with widths $>30^{\prime\prime}$. The Level 4 data also reveals the fine-scales structuring across the coronal hole, which is made evident in the Level 5 data.

\begin{figure}
\centering
    \includegraphics[scale=0.7]{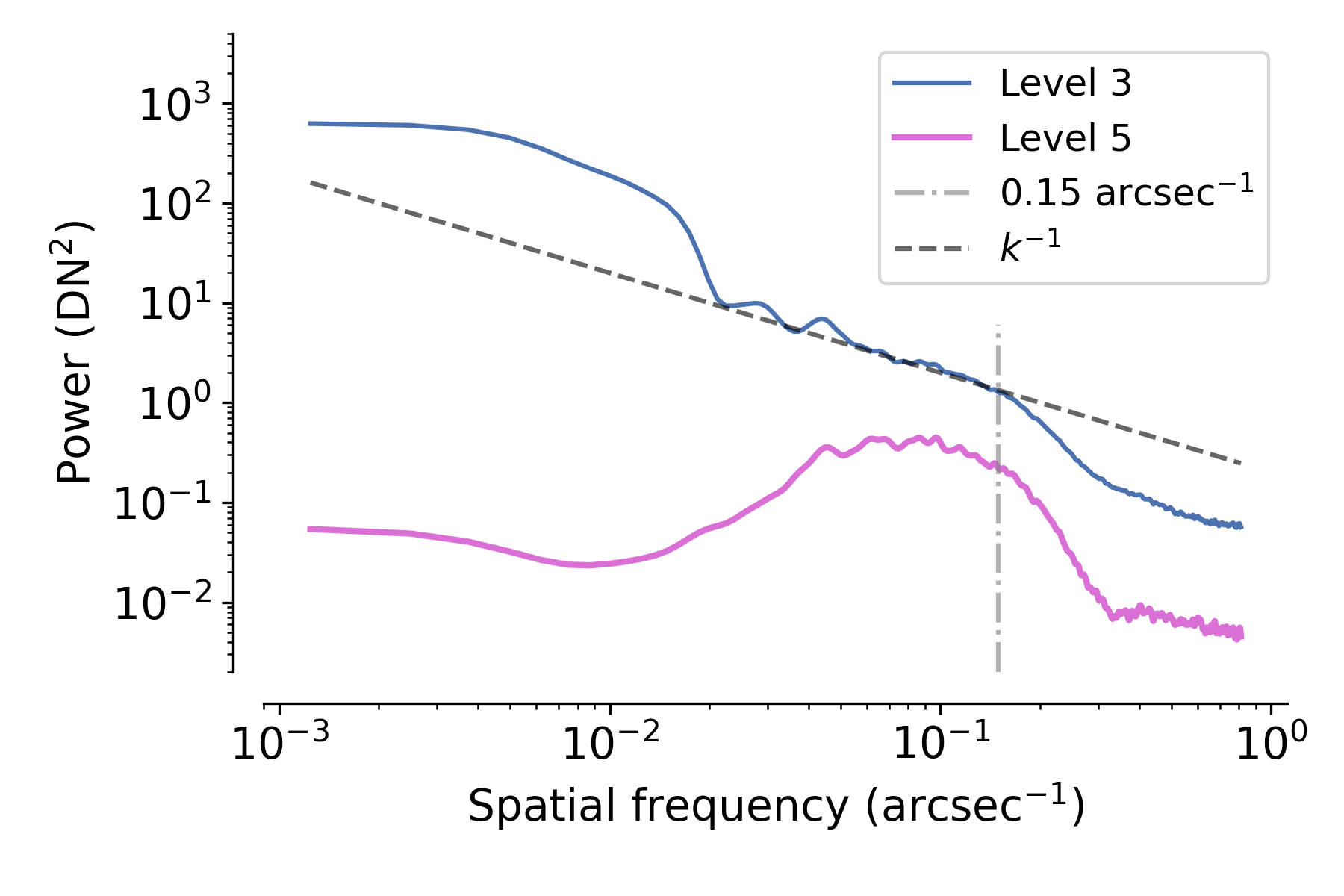}
    \caption{Discrete Fourier transform of Level 3 and 5 curves shown in 
    Figure~\ref{fig:int_curves}. The vertical dashed-dot curve marks the approximate position of the spectral break. The dashed line shows a $k^{-1}$ power law for comparison to the Level 3 data.}\label{fig:fourier}
\end{figure}

The intensity series from the level 5 data in Figure~\ref{fig:int_curves} also shows the filamentary structure has a weak signal, typically only a few DN above the local background emission at its brightest, which is around $<$1~\% of the total local emission. The typical spatial scale of these structures appears to be around 5-10$^{\prime\prime}$. This can be confirmed via a Fourier analysis of the spatial power spectrum, which is shown in Figure~\ref{fig:fourier}. The power spectra of the enhanced images clearly show distinct characteristic scales, with different frequency regimes displaying distinct power slopes separated by a spectral break \citep[also seen in the spatial power spectrum of equitorial plumes][]{Uritsky_2021}. The spectral break occurs around $\sim0.15$~arcsec$^{-1}$, shown by the vertical dashed line in the figure. Furthermore, we find that the spectral slope of the level 3 data is around $k^{-1}$ for spatial frequencies between $2\times10^{-2}-0.15$~~arcsec$^{-1}$.

\subsection{Variations with height}\label{sec:results_height}

Given the lack of overlapping emission features present in the coronal holes (as compared to on disk), there is the opportunity to examine how the fine-scale structure evolves as a function of height. To achieve this,
a number of individual features are selected and extracted from the image by taking slices perpendicular to the features axis, with slices separated by 1 pixel in height. The features chosen are the brightest features in the image, hence providing the highest signal to noise examples. We then fit a Gaussian function to the cross-sectional profile of the filamentary structures, performing this for each individual cross-sectional slice along the length of the structure. The Gaussian is modified with the addition of a linear function to account for any intensity gradients across the slice.

To fit the model to the cross-section we utilise a Bayesian approach. For the prior distributions we use Half-Normal distributions for the amplitude, $A$, and the width of the Gaussian, $\sigma$, and Normal distributions for the Gaussian centre, $\mu$, the constant, $C$ and the gradient of the linear function, $m$. For the likelihood we assume that the errors on the measured intensity are normally distributed about their true values. The model summary is then:
\begin{eqnarray}
    A&&\sim \mbox{HalfNormal}(10),\nonumber\\
    \mu &&\sim \mathcal{N}(0,3),\nonumber\\
    \sigma&&\sim \mbox{HalfNormal}(5),\nonumber\\
    C&&\sim\mathcal{N}(0,5),\nonumber\\
    m&&\sim\mathcal{N}(0,1),\nonumber\\
    f_i &&= A\exp\left[-(x_i-\mu)^2/(2\sigma^2)\right]+C+mx_i,\nonumber\\
    I_i &&\sim \mathcal{N}(f_i,\epsilon_i),\nonumber
\end{eqnarray}
where $I_i$ is the intensity at the spatial location $x_i$. Each intensity value has its own likelihood with standard deviation $\epsilon_i$ calculated from the expected uncertainties of the 
flux \citep{YUANAK2012}. The prior distributions are not uninformative, as is often seen in the solar literature \citep[e.g.,][]{2013A&amp;A...552A.138V,Pascoe_2017}. The priors were chosen to provide a degree 
of regularisation to the model, helping to fit the data for cross-sections where the emission is only just above the background. Such cases were found to be prone to diagnostic issues if the prior distributions for model parameters where uninformative. The chosen priors are still only weakly informative, providing little information to the posterior. Following the suggestions of \cite{Gabryetal2019}, prior predictive checks are used to confirm that the prior model only weakly identifies the observational model. 

\begin{figure}[!t]
\centering
    \includegraphics[scale=0.7]{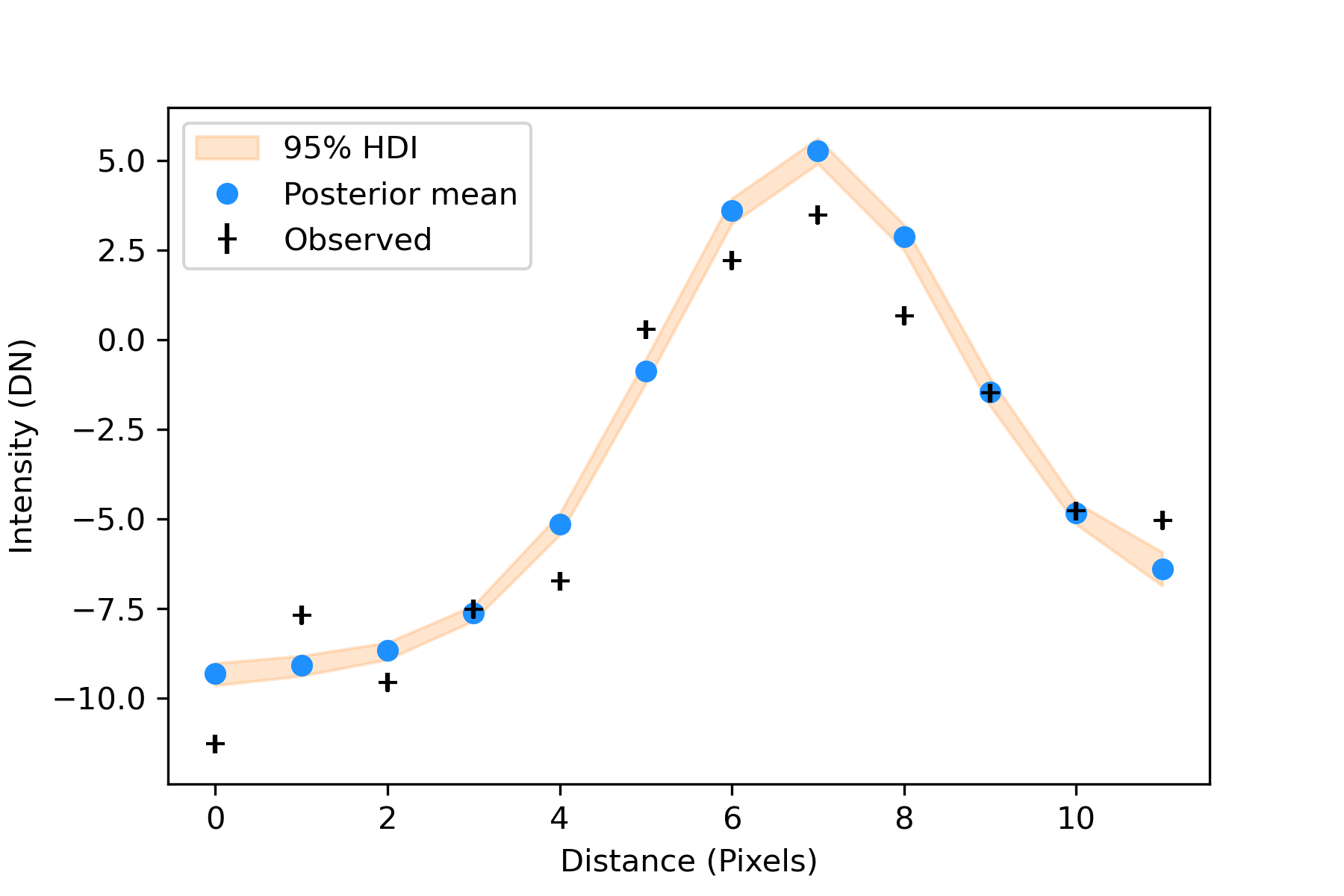}
    \caption{Example of Gaussian fit to cross-sectional intensity profile. The black crosses are
    the observed intensity values from the Level 5 data. The blue dots show the posterior
    means for the intensity and the peach band corresponds to the 95\% highest density interval (HDI).}
    \label{fig:bayes_exam}
\end{figure}

To sample the posterior we use a Hamiltonian Monte Carlo (HMC) sampler \citep[making use of PyMC3 -][]{pymc3}, running multiple independent sampling chains. The use of multiple chains enables us to use various diagnostic tools to assess model
convergence \citep[e.g., the $\hat{r}$,][]{bda_3rd} and HMC also naturally identifies problematic posteriors \citep{Betancourt2017}. A number
of fits to the cross-sectional profiles were found to have issues and were discarded from further analysis. An example of the model fit to a cross-section is shown in Figure~\ref{fig:bayes_exam}.

\begin{figure*}
    \includegraphics[scale=0.65]{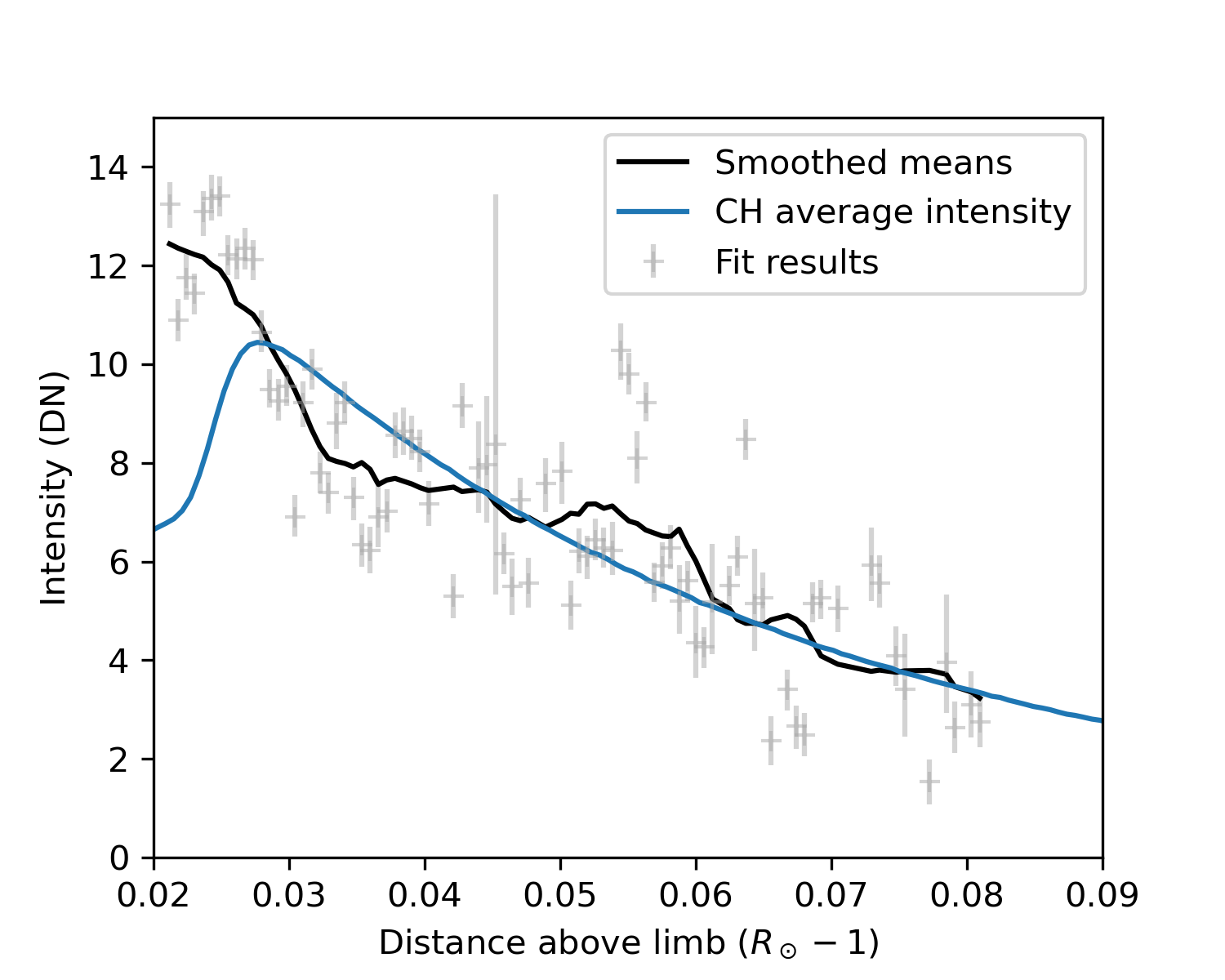}
    \includegraphics[scale=0.65]{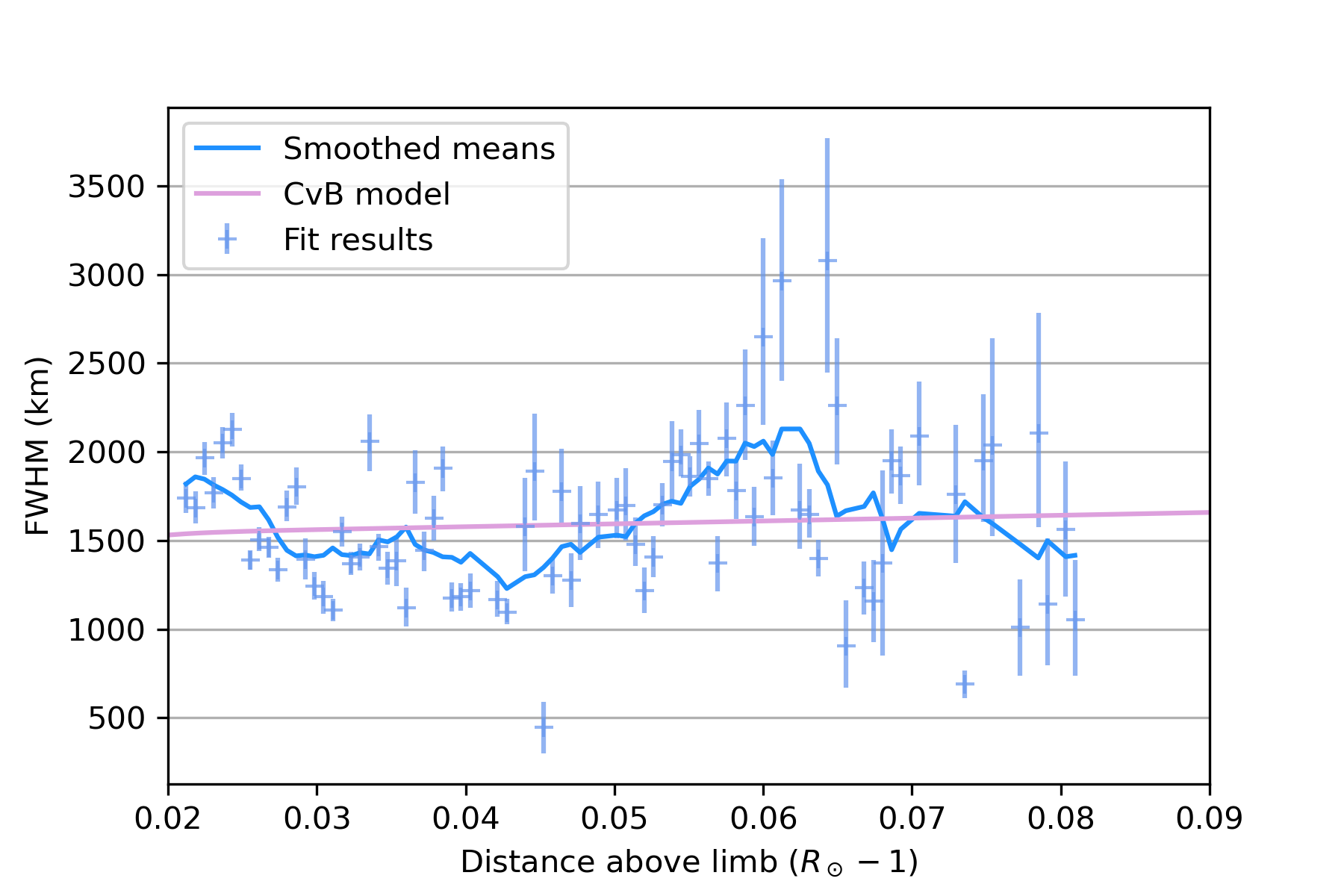}
    \includegraphics[scale=0.65]{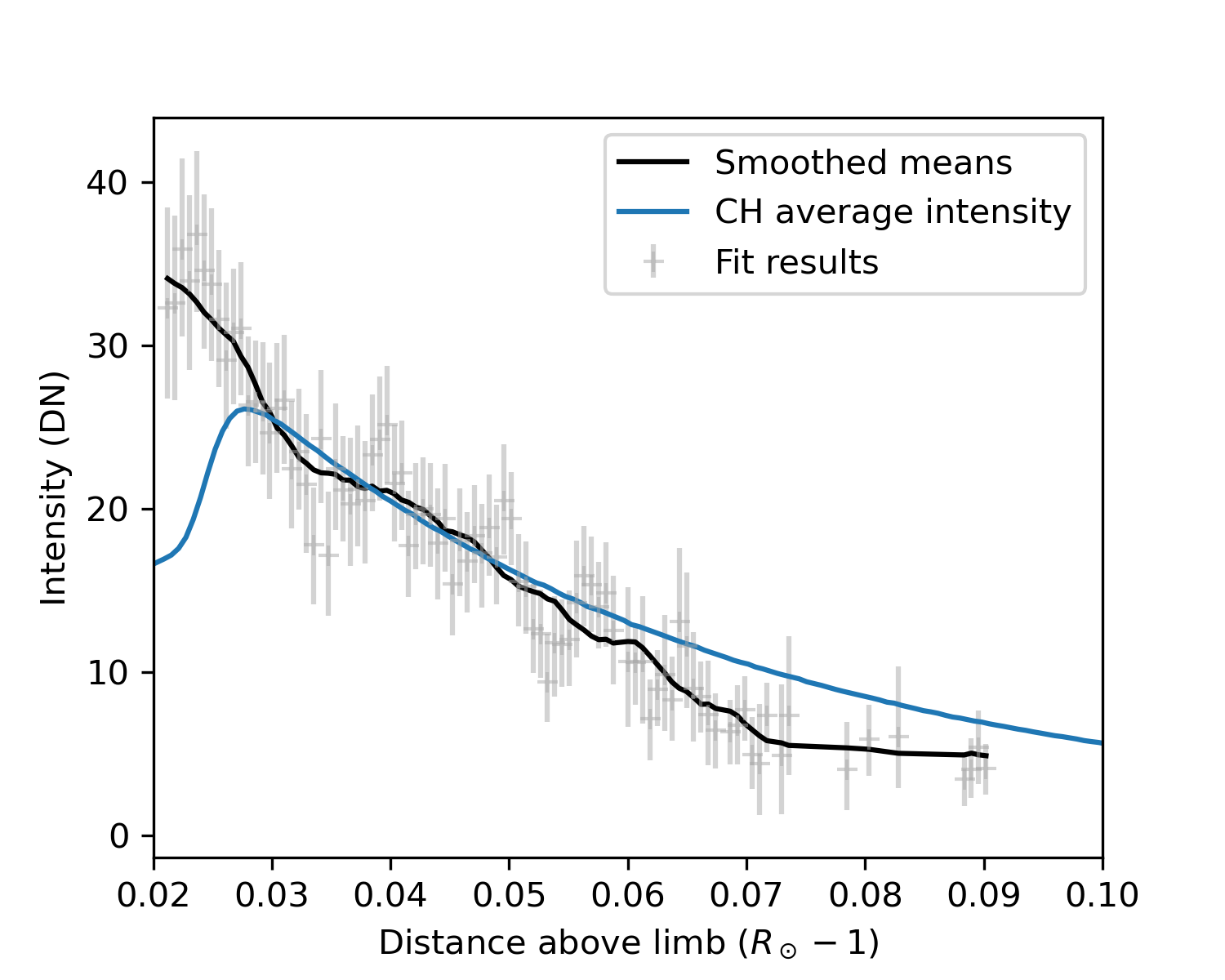}
    \includegraphics[scale=0.65]{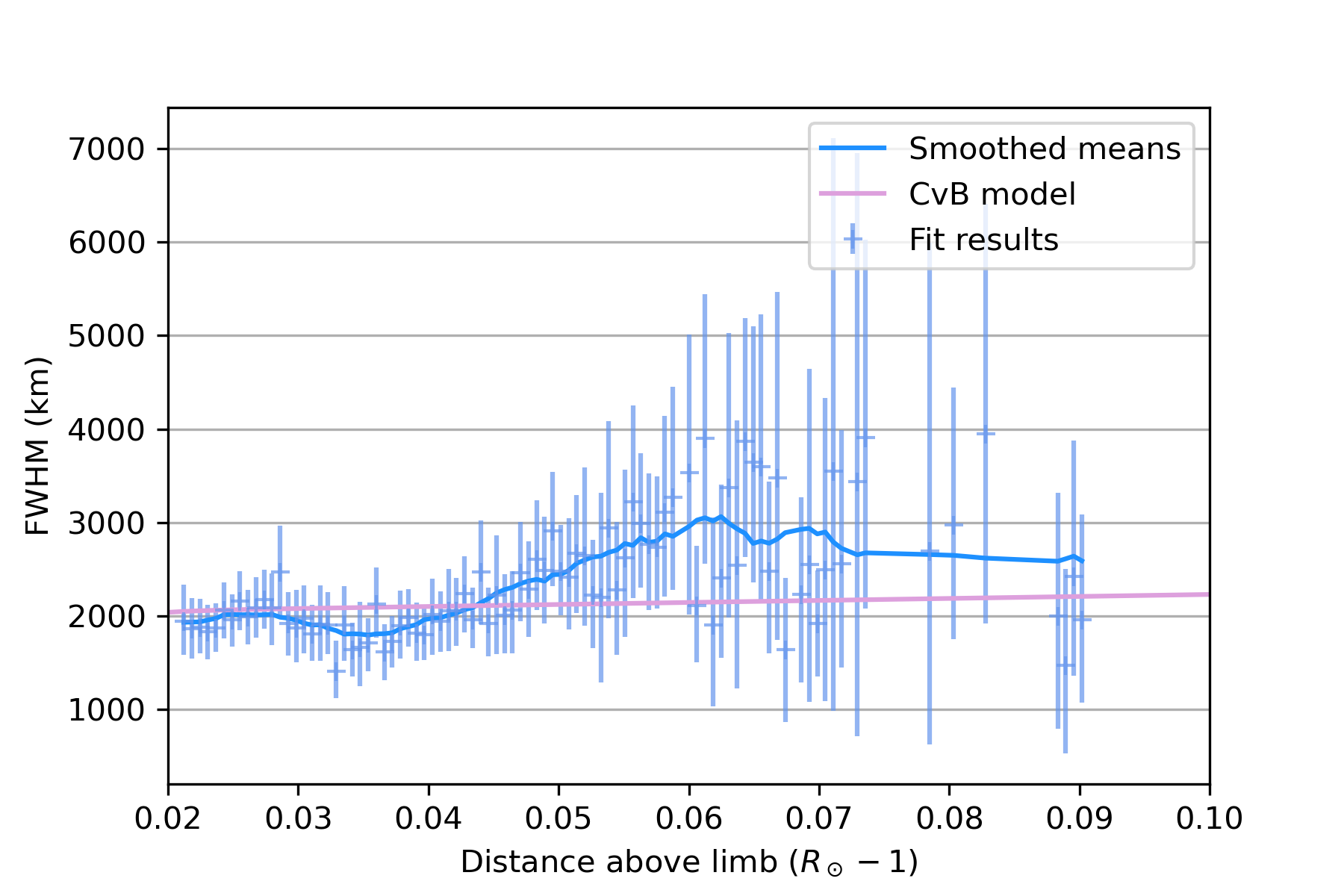}
    \includegraphics[scale=0.65]{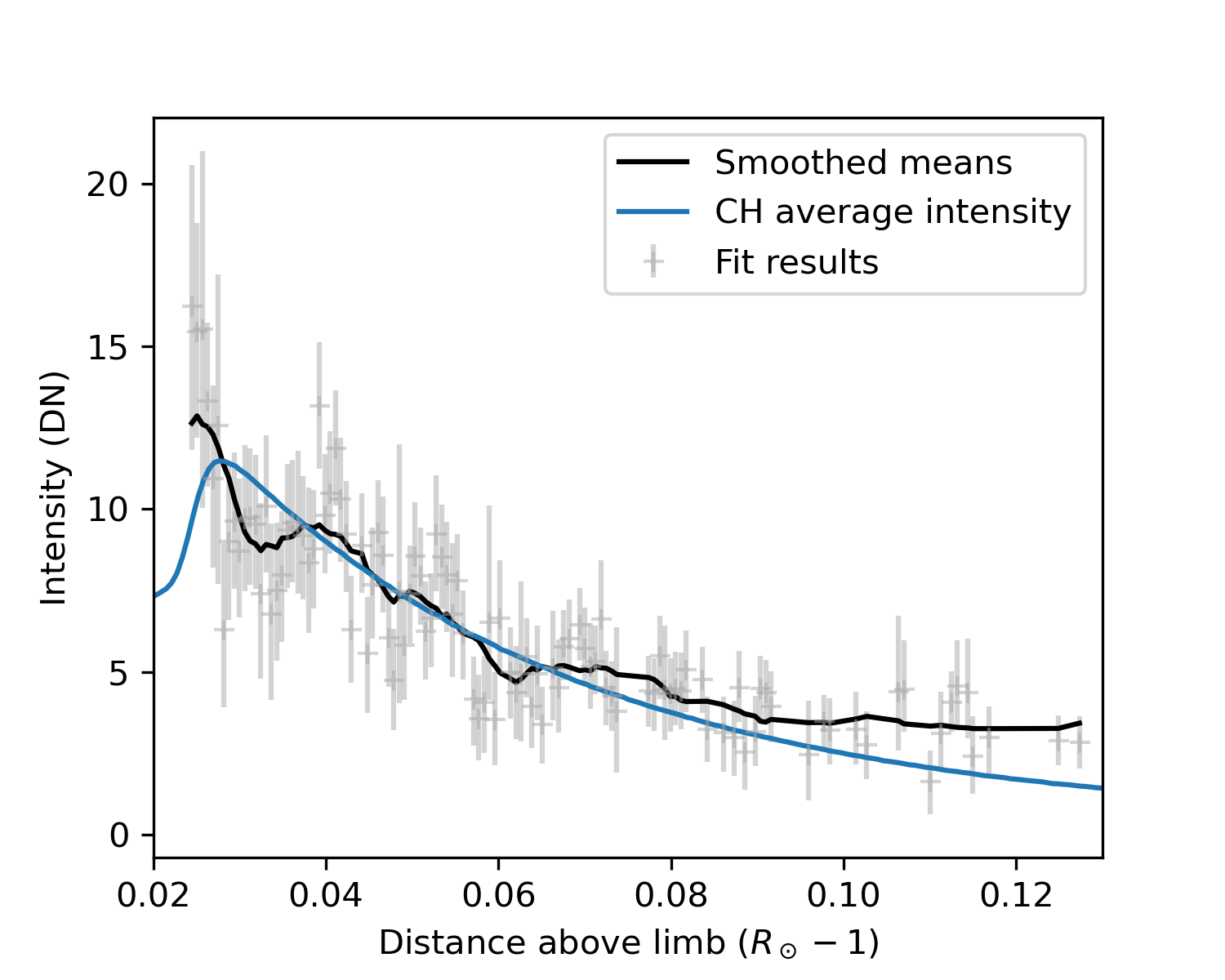}
    \includegraphics[scale=0.65]{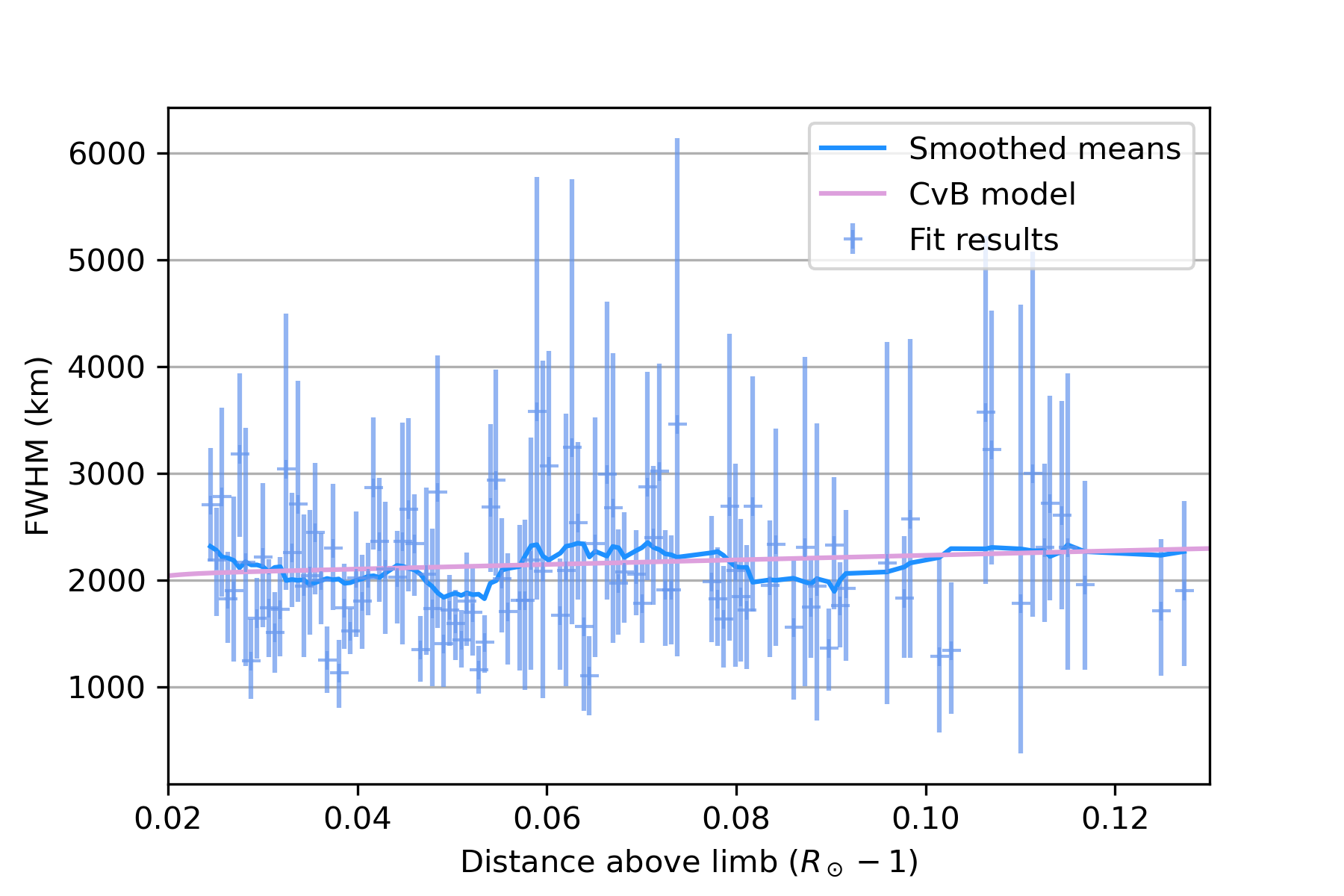}
    \caption{Results from the fit of a Gaussian function to the cross-sections of fine-structure observed in the coronal holes. The results are given for height above the limb in solar radii. In the text, these are discussed as plumes 1, 2, and 3, in descending order. The left hand plots show the peak intensity, with the posterior mean values for each fit shown as a grey cross, and the vertical length of the cross-arms represents the 94\% highest density interval for the posterior. The black solid line is the running average of the means and the blue line is the average intensity across the coronal hole. The right hand plots show the Gaussian Full-Width
    Half Maximum (FWHM). The posterior mean values are the blue crosses and the the vertical length of the cross-arms represents the 94\% highest density interval for the posterior. The blue solid line is the running average of the posterior means. The pink line show the expected expansion from the polar coronal hole magnetic field model in \cite{CRAVAN2005}.  }
    \label{fig:p13_vs_height}
\end{figure*}

\begin{figure*}

    \includegraphics[scale=0.65]{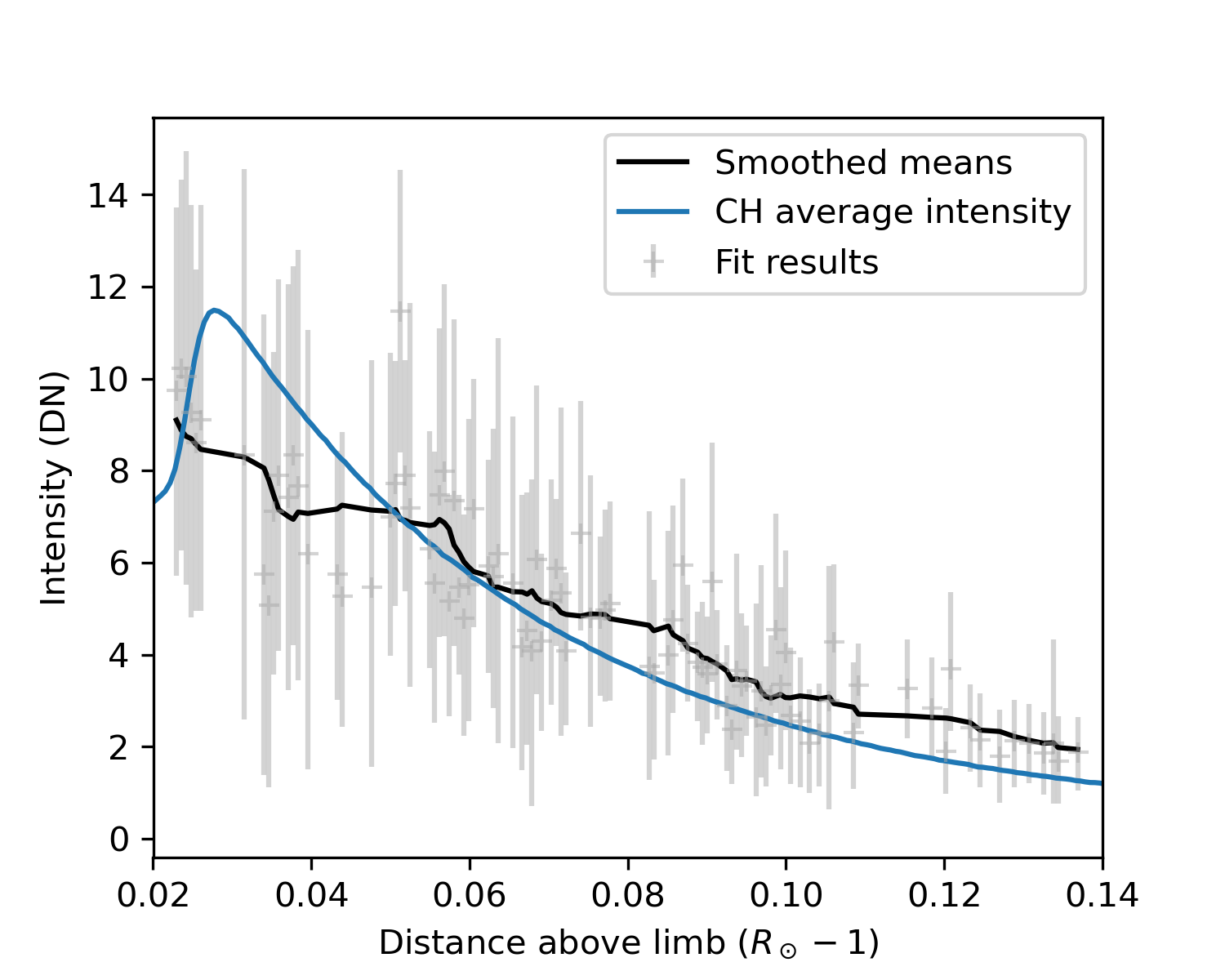}
    \includegraphics[scale=0.65]{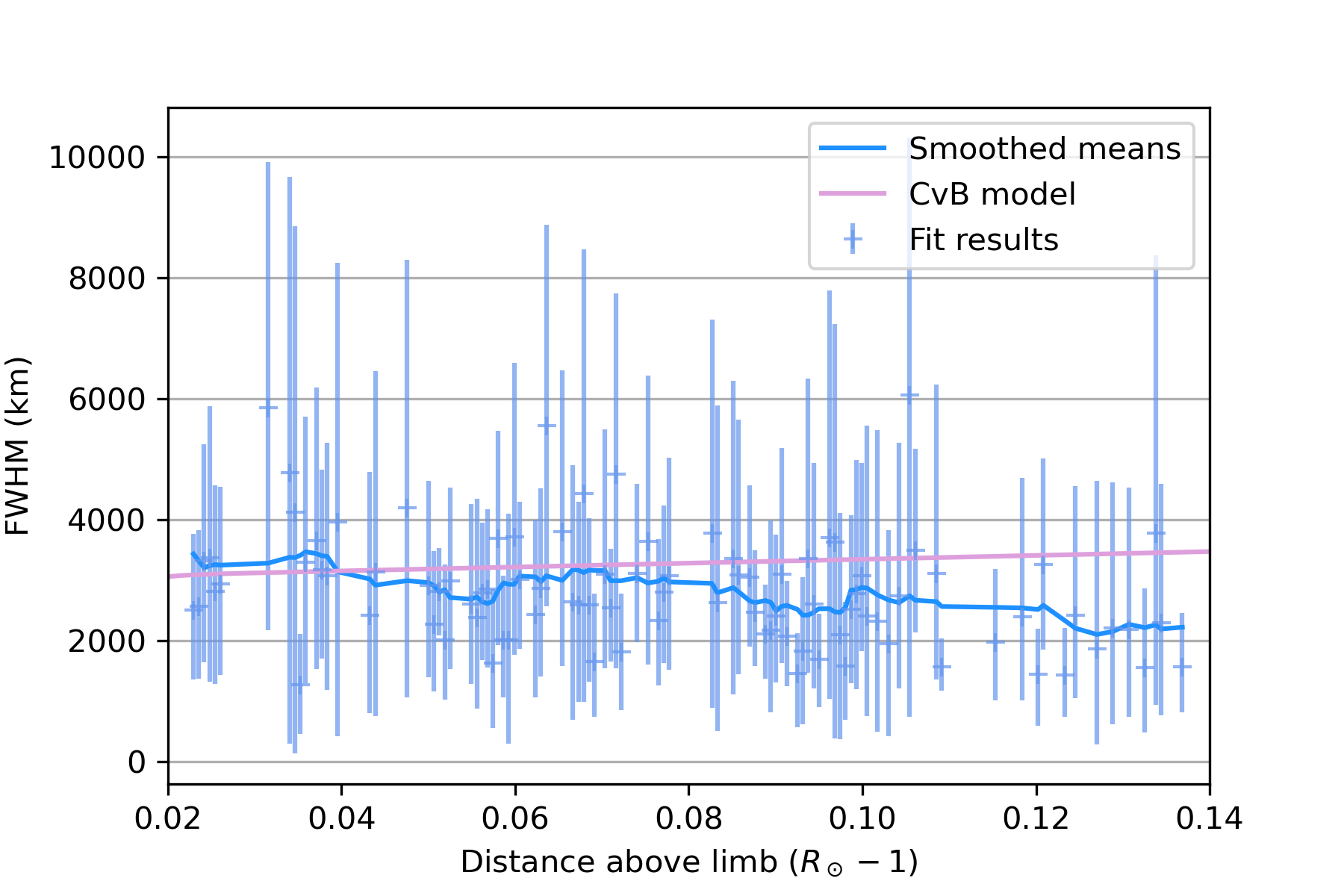}
    \includegraphics[scale=0.65]{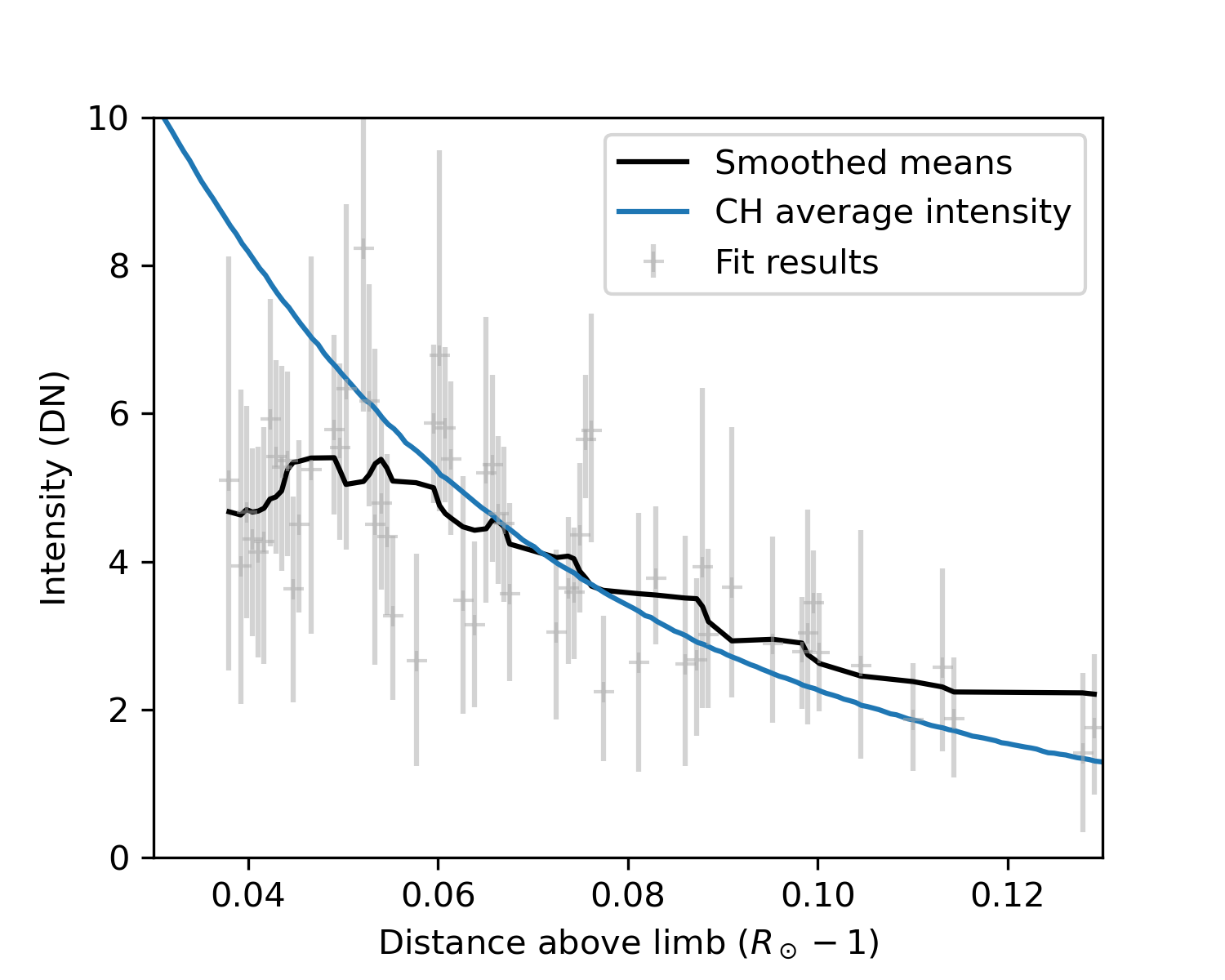}
    \includegraphics[scale=0.65]{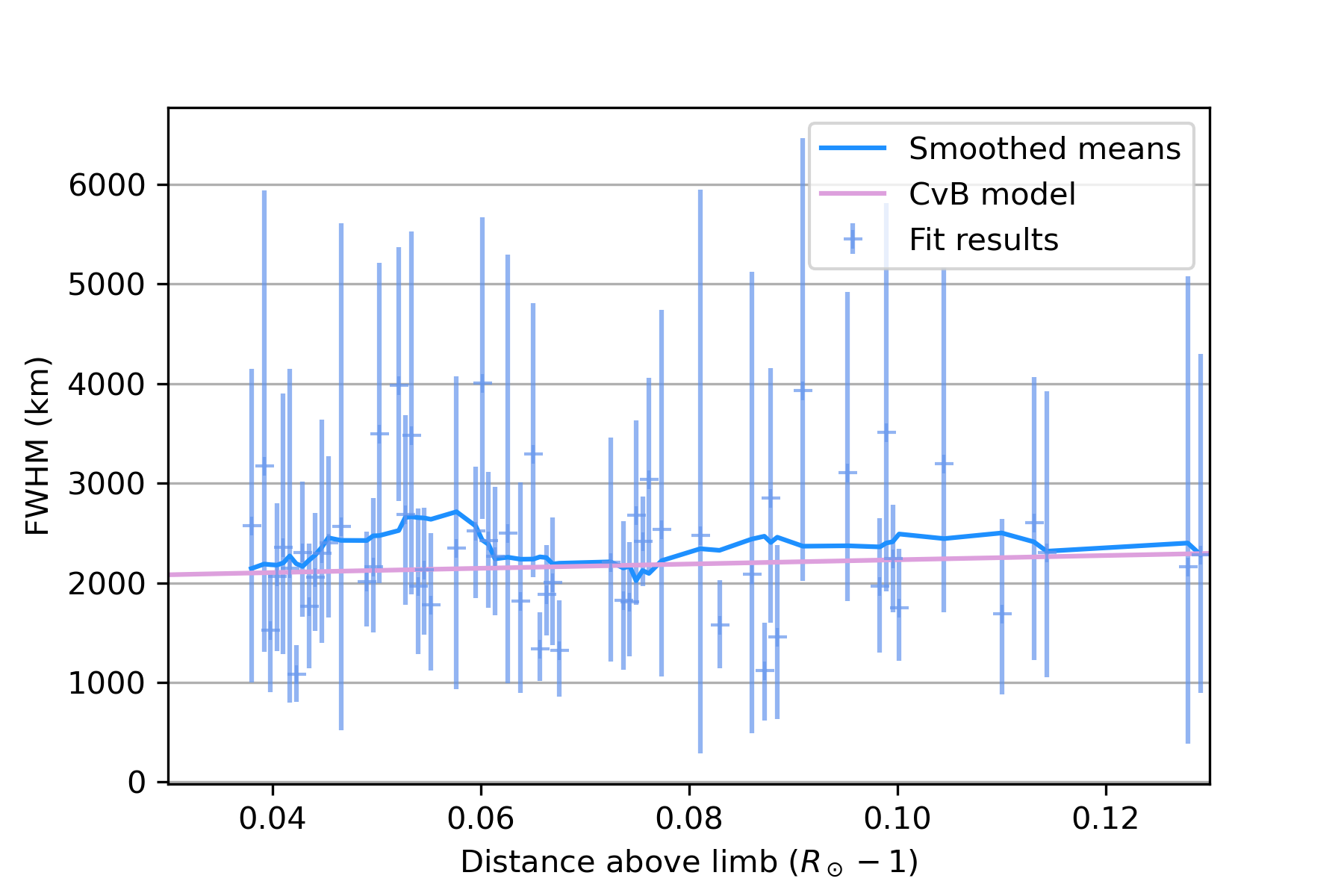}
    \caption{See Figure~\ref{fig:p13_vs_height} caption for details. In the text, these are discussed as plumes 4 and 5.}
    \label{fig:p45_vs_height}
\end{figure*}

\medskip
In the left-hand panels of Figures~\ref{fig:p13_vs_height} and \ref{fig:p45_vs_height}, we show the intensity values for a number of different fine structures as a function of height. The intensity values correspond to the amplitude of the fitted Gaussian ($A$). The marginal posterior mean and the 94\% HDI are given by crosses and error bars. As there is some variability in the measured values of intensity, the running mean is shown. As one might expect from the images, the intensity is found to decrease as a function of height.  The change in intensity along the individual features is consistent with the average decrease in intensity across the coronal hole, i.e. as measured from the Level 3 data. As a reference point, we also display the average off-limb intensity within the coronal hole. To permit a comparison, the average intensity is scaled to be close to the intensity of the fine-structure in each plot. 

In the right hand panels of Figures~\ref{fig:p13_vs_height} and \ref{fig:p45_vs_height} we display the width (full-width-half-maximum calculated from $\sigma$) of the fine-scale features. It can be seen that the uncertainties on the widths of the filamentary structures are variable as a function of height. The cross-sectional profiles closer to the limb are brighter and have a higher S/N, hence the uncertainties are smaller and on the order of $\sim100$~km for the $94\%$ highest density interval (HDI). With increasing height above the limb, the intensity of the structures drops and the S/N decreases. Hence, the uncertainties on the parameters in the Gaussian model increase and become reasonably large\footnote{It is worth noting that there could also be issues with model misspecification. The linear function used for the background might not be adequate in some cases.}. In each plot we also show the expected change in flux tube width to due to expansion \cite[based on the coronal hole model discussed in ][]{CRAVAN2005}. The change in width is calculated assuming there is conservation of magnetic flux within a flux tube that possesses a cylindrical cross-section. The change in width inferred from the magnetic field model is expected to be small, on the order of a pixel or so, over the observed height range. This makes it challenging to determine whether there is evidence for expansion of the fine-scale features with height. To discern any trends as function of height, we have overplotted a running mean for the values (calculated with a uniform window of $\sim9^{\prime\prime}$). The measured widths of the fine-structure are generally consistent with an increase in area due to magnetic field expansion. The running means of plumes 3 and 5 appear to follow the expected trend well, while plume 2 suggests greater expansion. For plume 1 a trend is not discernable, while for 4 the width appears to decrease with height. Hence, from these results it does not seem wise to draw definite conclusions about whether the filamentary structure expands with height.

\medskip

\subsection{Time-scales}
\begin{figure}
\centering
     \includegraphics[scale=0.7]{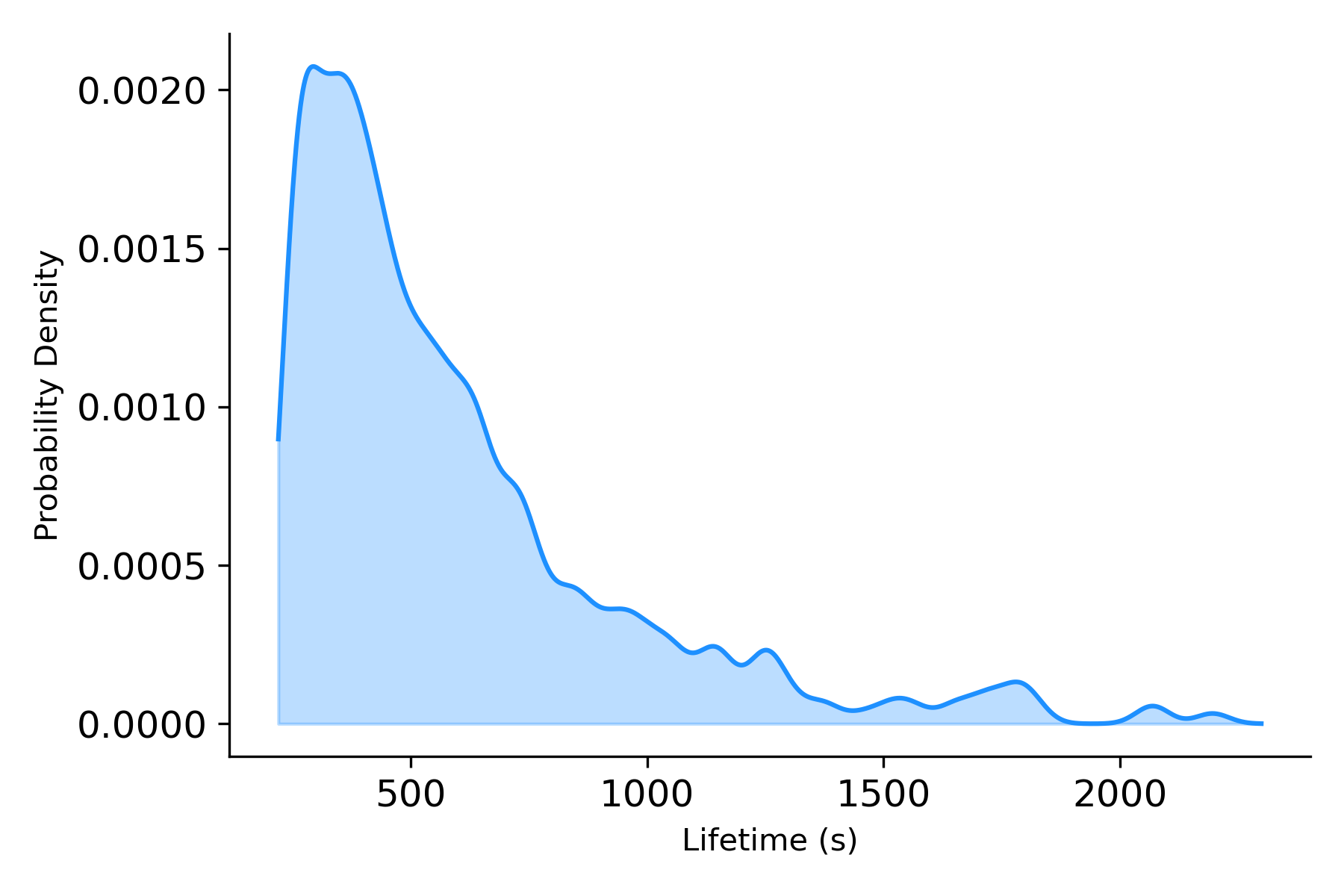}
     \caption{Probability density of the life-time of the fine-structure in coronal holes.}\label{fig:dur}
 \end{figure}

\textbf{It is of interest to estimate the time-scales associated with the fine-scale structure. In order to do this, we make use of a methodology the we have previously used for measuring transverse waves \citep{WEBetal2018,MORetal2019,Weberg_2020}. In short, the method finds significant intensity peaks in time-distance diagrams that are assumed to the be associated with emission from coronal structures. These intensity peaks are then linked together in time, subject to various conditions on their spatial and temporal locality with previous peaks. The result is many time-series that represent the central location of coronal structures in the time-distance diagram. We impose a lower limit on the duration of the time-series to reduce spurious detections. We take the length of these time-series as an indication of the lifetime of the fine-scale structure. The density distribution for the lifetimes is derived using kernel density estimation (with 5-fold cross-validation employed to select the bandwidth) and is shown in Figure~\ref{fig:dur}. The mean value for the lifetime is around 600~s.}

\section{Discussion}\label{sec:discussion}

Using a combination of data processing techniques we have been able to uncover fine-scale filamentary structure in polar coronal holes, which is largely hidden amongst the general diffuse emission. The structure is found to be present down to the smallest observable scales with AIA. The filamentary structure is near-radially oriented which would suggest that it is following the background magnetic field, and extends outwards towards the edge of the AIA field of view. This fine-structure likely extends into the middle corona \citep{Westetal2022}. Enhanced white light images from eclipse observations \citep{Druckm_ller_2014} appear to confirm this, revealing filamentary structure in polar regions. However, the scales of the features are unclear from these earlier results. Further, recent analysis of Solar Terrestrial Relations Observatory-A (STEREO-A)/COR2 data by \cite{DeForest_2018} shows that the fine-scale structures are still present out to $14~R_\odot$, where the corona transitions to the solar wind. As noted by \cite{DeForest_2018}, it should not be surprising that that the corona is finely structured given the dominance of magnetic pressure over gas pressure.

\medskip 

The fine-scale structuring we observe in AIA is potentially the micro-plumes/network-plumes that were hypothesised by \cite{Gabriel_2009}. The fine-scale structure is present across the
polar coronal hole, in both plume and inter-plume regions. However, \cite{Gabriel_2009} suggest that `\textit{[they are] dispersed in an interplume medium of the regular background corona}', which does not appear to be the case. The fine-scale structure is also found within the plumes, as hypothesised by \cite{Wilhelm_2011}. The fine-scale structure is also seen directly in on-disk plumes by \cite{Uritsky_2021}, which confirms it is not just confined to inter-plume regions.

\subsection{Over-density}
The intensity of an optically thin-emission line is given by
\begin{equation}
    I(\lambda)=A_X\int C(T, \lambda, n_e)n_e^2 dz,
    \label{eq:int}
\end{equation}
where $A_X$ is an abundance factor, $C$ is the contribution factor, $n_e$ is the electron density,
and this is integrated along the line of sight. For the Fe IX emission in the 171~{\AA} passband, the intensity is only weakly dependent on temperature hence the variations in intensity are largely
dictated by variations in density. This suggests that the observed fine-scale structure is over-dense compared to the surrounding plasma. The source of the over-density could potentially be mass-flows driven by the small-scale reconnection events discussed in \cite{Raouafi_2023}.  However, the small amount of additional emission from the fine-scale structure (leading to only a few DN signal on the sensor) implies the fine structure is only marginally over-dense compared to the ambient emission. 

The presence of weakly over-dense structures is also supported by the observations of undamped kink waves in coronal holes from 
\cite{MORetal2015} and \cite{Weberg_2020}. In the presence of a continuous density profile across a wave guide, the kink mode is expected to experience damping due to resonant mode conversion \citep[demonstrated numerically for propagating modes by, e.g.,][]{PASetal2011}. The rate of resonant damping is controlled by the density difference between the wave guide and surrounding medium \citep{TERetal2010c}. Hence, the observation of undamped kink waves implies a small density contrast. Damped propagating kink waves are observed in the quiet Sun although this damping is only weak, again indicating marginally over-dense plasma structures in the quiet Sun \citep{VERTHetal2010,TIWARI_2019,tiwari_2021,Morton_2021}.  Even in active region loops the over density appears to be small, i.e., on the order of $\sim2-3$ \citep{Asensio_Ramos_2013,Pascoe_2018}. 

\cite{Gabriel_2009} suggested that `micro-plume' over-density is factor of 4-5, which would not agree with the inferences from wave studies. A further study is required to determine the temperature and electron density of these fine-scale features, something which may be possible with AIA but might require the high-resolution coronal observations afforded by DKIST \cite[e.g., Cryo-NIRSP ][]{Fehlmann_2023}.

\subsection{Spatial and temporal structure}
Without observations of the middle corona, it is not a foregone conclusion that the structuring observed in the AIA images presented here is related to the fine structure observed in the upper corona. However, we believe it to be the case. The measured size of the structures in the low corona is not at odds with the size of the structures found in the upper corona.  \cite{DeForest_2018} reports structures down to the size of $\sim$20~Mm at 10~$R_\odot$, and suggest they should correspond to structures on the scales of 300~km at the chromospheric level, if their expansion follows the super-radial expansion of plumes. Assuming a 300~km scale at the chromospheric level and accounting for the observed rapid expansion of plumes up to height of 20-30~Mm above the surface ($\sim1.029-1.43$~$R_\odot$), this would suggest structures on the order of 1500-3000~km in the low corona. These values are in line with the widths measured for individual structures here. 

\cite{DeForest_2018} note the origin of the fine-structure is unclear but highlight the potential role of photospheric dynamics via the magnetic carpet or the intrinsic dynamics of the corona could be responsible for its generation. Given the structure is present at the base of the corona, it would seem that the coronal dynamics are not responsible. We recall that small-scale jets are found to be related to enhancements in the plume emission \citep{Samanta_2015}. IRIS observations also revealed the prevalence of jets at the network boundaries with widths of $\le 300$~km \citep{Tian_2014}, thought to be related to spicules which also have similar widths \citep{PERetal2012}. Hence, the fine-scale structure observed throughout the corona in polar regions could be the result of mass-loading of magnetic field lines from these small-scale jets. There is also the reconnection driven jets discussed by \cite{Kumar_2022} and \cite{Raouafi_2023}, which appear less frequently than spicules/network jets but are more energetic. The occurrence of both of these jets is thought to depend on the photospheric dynamics \citep[e.g.,][]{MARetal2017}. \textbf{\cite{Raouafi_2023} suggest that the lifetime of the small-scale coronals jets are around 5-10 minutes, which is line with the estimated lifetimes of fine-scale structures observed here (Figure~\ref{fig:dur}).}

\medskip

\begin{figure*}[ht!]
    \includegraphics[scale=0.7]{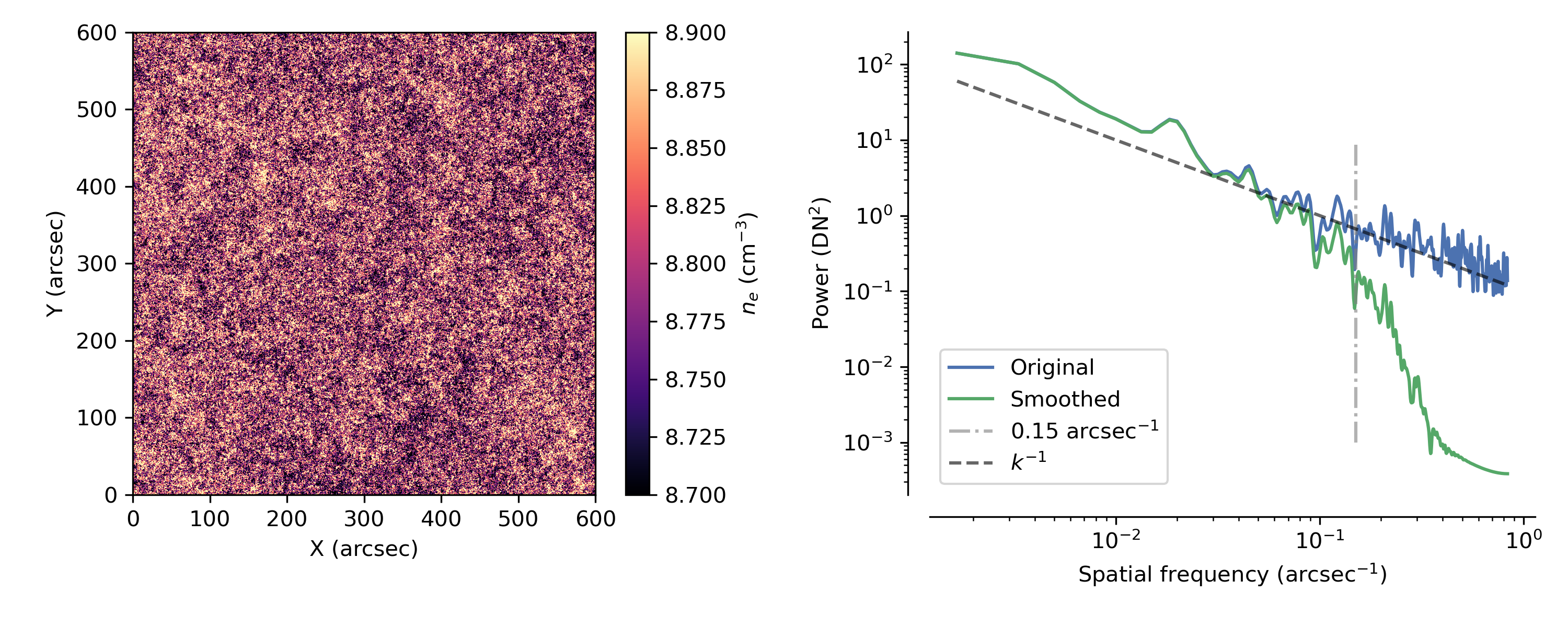}
    \caption{The spatial analysis of synthetic coronal emission. The left-hand image shows the density distribution through a coronal slab, which follows an isotropic $1/k$ scaling law. The colour scale has been clipped to the given values. The right-hand panel shows the analysis of the one-dimensional emergent intensity that is calculated from the coronal slab. The Fourier spectra of the original and smoothed emergent intensity are displayed, along with a curve indicating the slope of a $1/k$ power law.  } \label{fig:sim}
\end{figure*}

In Section~\ref{sec:results_perp} we demonstrate that the fine-scale structure has a spatial distribution that follows a spectral slope of $k^{-1}$. It is not evident that the underlying distribution of density structures would also follow the same scaling law. The emergent intensity depends on the contribution function and line-of-sight integration through the corona (see Eq.~\ref{eq:int}), which could modify the scaling law associated with the density structures. We undertake a simple simulation to examine whether the observed scaling law reflects the scaling law of the density structures. We begin by defining a two dimensional slab of coronal plasma that contains density structures distributed according to an isotropic power law structure of $k^{-1}$ (Figure~\ref{fig:sim} shows the electron density across the slab). The temperature of the coronal plasma is taken as proportional to the inverse of the density, which assumes that the magnetic field is constant throughout the slab. The density is chosen such that the mean value is $\log n_e=8.8$, and the standard deviation is 0.1, which is in line with previous measurements of density values in plumes and interplume regions \citep[e.g.,][]{Banerjee_2009}. The temperature scaling is chosen such that the temperature has a mean value of 1~MK. Then the one dimensional emergent intensity from the slab is calculated for Fe IX 171~{\AA} under the assumption of optically thin-emission (i.e., Eq~\ref{eq:int}), taking into account the response function of AIA \citep[calculations performed using FOMO][]{Van_Doorsselaere_2016}. 

The Fourier power spectra for the synthetic intensity is shown in the right panel of Figure~\ref{fig:sim}, which is found to follow a $k^{-1}$ power law, hence reflects the underlying density structure. We note that the overall magnitude of emergent intensity is larger in the  synthetic data than that observed in the actual data. This is likely because in the simulation we have integrated through a large slab of coronal plasma with the same mean density throughout. This does not take into account the variation in density expected along a ray. The variation arises due to the fact that a ray passes through coronal plasma at various heights above the photosphere. It should be expected that the emergent intensity is over-estimated in our simulation, around a factor of 4 or so.

The power law for the synthetic intensity continues to the smallest scales, which is not what is observed in Figure~\ref{fig:fourier}. However, the spatial resolution for AIA is thought to be on the order of 1.2-1.5$\arcsec$. Smoothing the synthetic intensity with a Gaussian filter (standard deviation of 0.6$\arcsec$) should provide an appropriate level of blurring. Performing this smoothing leads to a similar drop off in power below $0.15$~arcsec$^{-1}$ seen in the observed spectrum (Figure~\ref{fig:fourier}). This result may indicate that the observed spectrum for the density structure continues to smaller scales.

It is worth noting that the simulated density structure does not contain features that look like the `beam' plumes, which are found to originate from strong unipolar magnetic fields in the photosphere \citep{DeForest_1997}. The concentrations of magnetic field are likely located at the boundaries of supergranule cells, where flux is found to accumulate \citep[e.g.,][]{Zirin_1985,GOSetal2014}. In the observed spatial power spectrum there is a jump in power at $\sim2\times10^{-2}$~arcsec$^{-1}$ (Figure~\ref{fig:fourier}), roughly at the supergranule scales ($\sim30$~Mm). Hence, super-granules may control the structuring of the corona above this spatial scale. A more detailed simulation would require some inclusion of the influence of network structure \citep[e.g., ][]{Gabriel_2009}.

\subsection{Implications}
The presence of fine-scale structuring of the coronal plasma has implications for understanding a broad range of phenomenon in the Sun's atmosphere. The influence of spatial inhomogeneities across magnetic field lines has already been discussed in some detail by \cite{DeForest_2018}, who highlighted the potential impact for understanding the solar wind, reconnection, the origin of solar wind turbulence and the nature of the Alfv\'en surface.  

It is worth realising that the presence of the spatial inhomogeneities also has implications for MHD wave propagation through the coronal holes (as well as the rest of the corona). It is known that the presence of density variations perpendicular to the magnetic field leads to a coupling of the MHD modes \citep{2019ApJ...873...56M}. Of particular note, especially with relevance to Alfv\'enic waves, is that the inhomogeneities in density ensure the presence of surface Alfv\'en waves \citep{GOOetal2012} rather than the pure Alfv\'en waves present in a plasma that is homogenous perpendicular to the magnetic field. The magnetic fields or flux surfaces are no longer independent and do not oscillate independently. In this scenario, resonant mode conversion is able to occur, along with an associated phase mixing \citep{SOLetal2015}. Furthermore, the inhomogeneity could also permit the development of instabilities, in particular the Kelvin-Helmholtz instability \citep{TERetal2008b,antolin2019}, as well a parametric instability \citep{Hillier_2018}. 

Many wave-based models of the heating and acceleration of the plasma in coronal holes typically assume the plasma is homogenous perpendicular to the field \citep[e.g.,][]{CRAVAN2005,SUZINU2005,VERDetal2012,van_Ballegooijen_2016,Shoda_2018}. The consequence of this is that the additional mechanisms for mode conversion and dissipation are missing from such models, which would influence the plasma heating and mass loss. Numerical simulations suggest that the Alfv\'enic waves in the low corona (specifically in the sub-sonic region up to 4~$R_\odot$) are expected to be close to linear \citep{Matsumoto_2012}, and this is also inferred from observational estimates \citep[e.g.,][]{THUetal2014,MORetal2015,WEBetal2018,Weberg_2020} where observed waves amplitudes in coronal holes are $\sim15-30$~km/s with propagation speeds of 300-400~km/s (hence, $\delta v/v_A\sim0.1$). The linearity of waves inhibits shock formation and also suppresses any turbulent cascade. Hence, phase mixing, resonances and instabilities may play a proportionally greater role in plasma heating at these lower heights than else where in the coronal holes.

Furthermore, \cite{Magyar_2022} provide a demonstration that the structure of the density inhomogeneity can also impart itself upon the perpendicular spectrum of the Alfv\'enic waves. In the simulations of \cite{Magyar_2022}, the distribution of density inhomogeneities are described by a power law and the perpendicular spectrum of the propagating waves evolves towards the same scaling. This is due to resonant absorption and phase mixing of the surface Alfv\'en waves. Hence, this process could explain the observed $1/f$ spectrum of the Alfv\'enic waves in the energy-containing scales found in the solar wind.

\section{Conclusion}
The solar corona appears to possess fine-scale density structures throughout, from the low corona out to where the corona transitions to the solar wind ($\sim$14~$R_\odot$). At least in coronal holes, there is growing evidence that its presence is due to reconnection-driven jets which are able to mass-load the magnetic field, e.g., \cite{Kumar_2022,Raouafi_2023}. For various reasons, most investigations into the coronal fine structure has focused on the scales associated with bright coronal loops in active regions e.g., \cite{ASCNIG2005,BROetal2012,BROetal2013,PETetal2013,antolin2014,Williams_2020}. Considerably less attention has been given to the fine structure of the corona in the quiet Sun and coronal holes. This may be due in part to the difficulty in revealing the fine-scale structure, which requires careful processing of the data to reveal its presence. There is opportunity to probe the plasma properties of the fine-structure further with AIA and will likely be the subject of future work. Improved measurements of the spatial scales will hopefully be possible with the EUI \citep{Rochus_2020} instrument during the perihelions of the Solar Orbiter mission. Furthermore, the data due to arrive from DKIST's Cryo-NIRSP should also be a promising avenue for probing the finely structured corona. Although better targets might be equatorial coronal holes or other open field regions. Previous observations from the Coronal Multi-Channel Polarimeter have demonstrated that there is low signal from Fe XIII emission lines in the polar coronal holes which might make measurements challenging.

\section{Acknowledgements}

RJM is supported by a UKRI Future Leader Fellowship (RiPSAW—MR/T019891/1). We would like to thank
C. DeForest, P. Antolin and N. Magyar for assistance and valuable discussions. Data used has been provided courtesy of NASA/SDO and the AIA, EVE, and HMI science teams. It is freely available at \url{http://jsoc.stanford.edu/}. For the purpose of open access, the author(s) has applied a Creative Commons Attribution (CC BY) licence to any Author Accepted Manuscript version arising.

\section{Software}
{Data analysis has been undertaken with the help of NumPy \citep{Numpy}, 
matplotlib \citep{Matplotlib}, IPython \citep{IPython}, Sunpy \citep{sunpy}, Astropy \citep{astropy}, SciPy \citep{Scipy}, Scikit Image \citep{scikit-image}, PyMC3 \citep{pymc3}, SciKit Learn \cite{scikit_learn}.}

\appendix

\section{Estimates of noise levels}
Given the significant level of data processing required to get to Level 5 data, we provide an estimate of the data noise each stage of processing. It is not straightforward to estimate the noise in the data. However, we use the method suggested by \cite{OLS1993}.  For each level of data, we smooth the image with a uniform filter of spatial extent 1.8" square. The filtered image is subtracted from the original to provide an unsharp masked image (in an attempt to remove the large-scale image structure). The the standard deviation is estimated from the unsharp masked data and divided by the standard deviation of the filtered image to given the noise to signal ratio. For the Level 5 data we divide by the standard deviation of the filtered Level 3 data, which contains the full signal. 

\begin{figure}
    \includegraphics{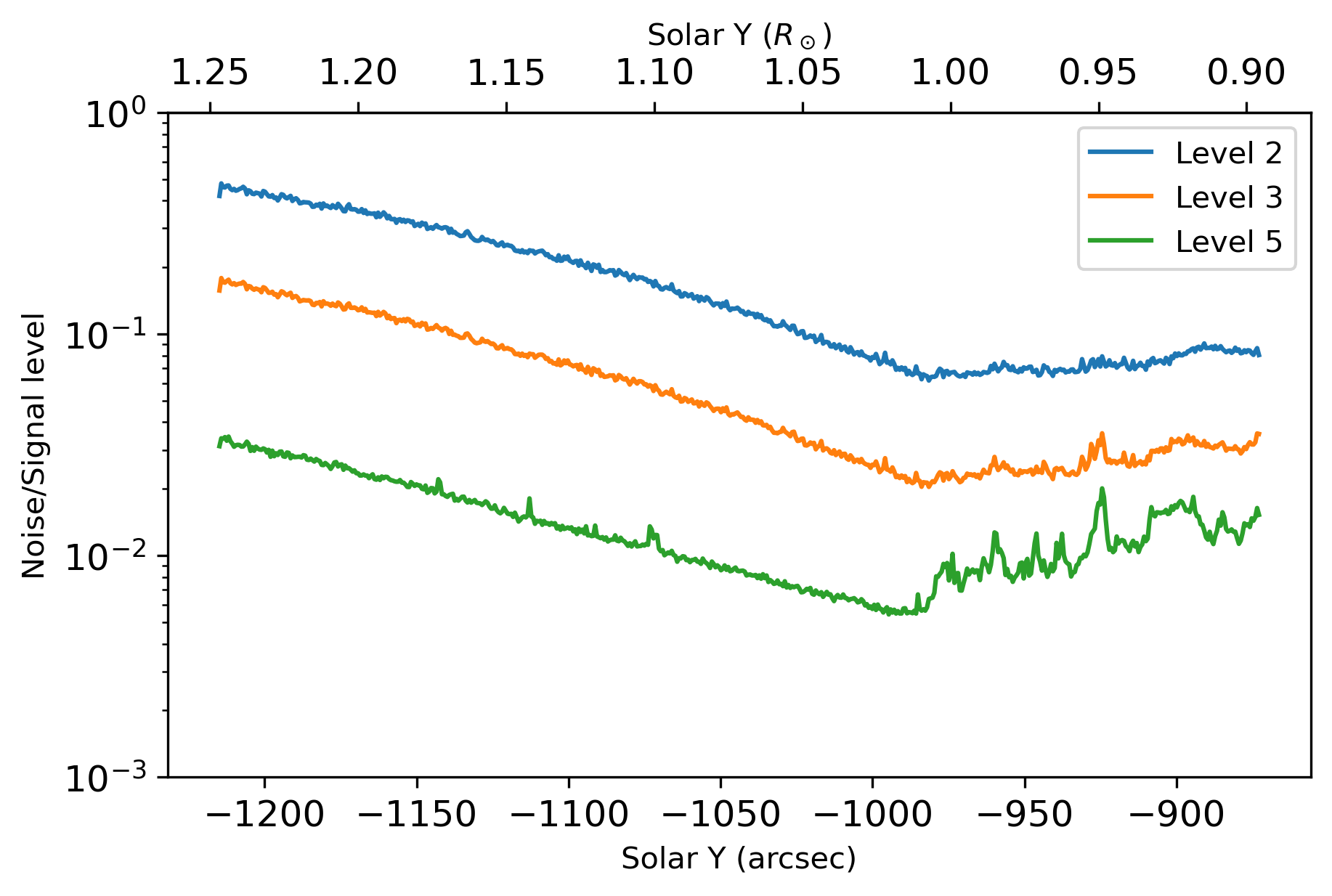}
    \caption{Estimate of image noise to signal level as function of distance from Sun center for the deconvolved (Level 2), noise-gated (Level 3) and unsharp mask averaged data (Level 5).}\label{fig:noise}
\end{figure}

We estimate the standard deviation as a function of pixel row in the images (Figure~\ref{fig:noise}) to demonstrate the increase in noise to signal level when moving away from the limb. Above the limb, between 
1-1.15~$R_\odot$, the noise is $\lesssim 1$~\% of the signal. This means the features we are measuring have signal above the image noise level.

We note that this unsharp mask method cannot remove all of the physical signal in the images. This problem is most pronounced on the disk which is why the noise to signal level is apparently much higher than the off-limb coronal hole. Hence, we expect the results shown in Figure~\ref{fig:noise} to be conservative estimates for the noise levels.

\bibliographystyle{aasjournal}
%\bibliography{mybib}

\end{document}